\documentclass[useAMS,usenatbib]{mn2e}
\usepackage{multicol}
\usepackage{graphicx}
\usepackage{epstopdf}
\usepackage{setspace}
\usepackage{changepage}
\usepackage[normalem]{ulem}
\usepackage{fmtcount}
\def\mbh{$M_{\rm BH}$}

\def\mbhlit{$M_{\rm BH,Lit}$}
\def\xmm{{\it XMM--Newton}}

\def\ergs{erg s$^{-1}$}
\def\gam{$\Gamma$}
\def\msun{$M_{\odot}$}
\def\comp{Comptonization}
\def\norm{$N_{\mathrm{BMC}}$}
\def\lx{$L_{\rm X}$}

\title{Measuring the black hole mass in Ultraluminous X-ray Sources with the X-ray Scaling Method}
\author[I. Jang et al.]{I. Jang,$^{1}$\thanks{E-mail: ij13@nyu.edu},
M. Gliozzi,$^{1}$ S. Satyapal,$^{1}$ and L. Titarchuk$^{1}$\\
$^{1}$ The Department of Physics \& Astronomy,
George Mason University, 4400 University Drive, Fairfax, VA 22030}

\begin{document}
\maketitle

\label{firstpage}
\begin{abstract}
In our recent work, we demonstrated that a novel X-ray scaling method, originally introduced for Galactic black holes, could be reliably extended to estimate the mass of supermassive black holes accreting at moderate to high level. 
Here, we apply this X-ray scaling method to ultraluminous X-ray sources (ULXs) to constrain their \mbh. 
Using 49 ULXs with multiple \xmm\ observations, we infer that ULXs host both stellar mass BHs and intermediate mass BHs. 
The majority of the sources of our sample seem to be consistent with the hypothesis of highly accreting massive stellar BHs with $M_{\rm BH}\sim100\,M_{\odot}$. 
Our results are in general agreement with the \mbh\ values obtained with alternative methods, including model-independent variability methods. 
This suggests that the X-ray scaling method is an actual scale-independent method that can be applied to all BH systems accreting at moderate-high rate. 
\end{abstract}


\section{Introduction}
It is now widely accepted that black holes (BHs) exist on very different scales, from stellar mass black holes (sMBHs) also known as Galactic black holes (GBHs) with the mass range of $M_{\rm BH}=3-20$ \msun\ to supermassive black holes (SMBHs) in active galactic nuclei (AGNs) or at the center of galaxies with the mass range of $10^6-10^9$ \msun, and possibly intermediate mass black holes (IMBHs) in ultra luminous X-ray sources (ULXs) and globular clusters with a mass range of $10^2-10^5$ \msun. 

ULXs are bright, off-nuclear X-ray sources whose X-ray luminosity (\lx) is larger than the Eddington limit for $10$ \msun\ BH (\lx\ $\ge10^{39}$ \ergs) \citep{roberts2007, FR11}. 
The nature of ULXs, specifically the BH class (sMBH vs. IMBH), is still debated. 
For example, ULXs may host sMBHs that are accreting at super-Eddington rate \citep{begel02,king09}, or their high luminosity can be  explained by anisotropic emission produced by relativistic beaming in sMBHs accreting at a normal rate, or by a combination of high accretion and beaming effects \citep[e.g.,][]{king2001, walton2011}. 
Another possible interpretation of ULXs is that they host IMBHs accreting at sub-Eddington rates and producing quasi-isotropic \lx\ \citep{CM99,maki00,kaa03}. 
Although there is no clear consensus about the nature of ULXs, it is possible that this class comprises both sMBHs and IMBHs. 
Three ULXs, M82 X-2, NGC 5907 ULX, NGC 7793 P13, have been confirmed to be a pulsar, reinforcing the likely heterogeneous nature of the ULX class  \citep{bachetti2014, israel16,israel17a, israel17b}.
The study of ULXs may therefore yield crucial information to further our understanding of the accretion process around BHs. 
For instance, ULXs harboring highly accreting sMBHs may shed some light on the accretion physics at near or super-Eddington rate \citep{ohsu05}. 
On the other hand, ULXs with IMBHs may play an important role in understanding the formation of the seeds of SMBHs in the early universe \citep{ebis01,volo10}.



Until now, no direct dynamical method has been successfully applied to ULXs due to the limited information about their optical counterpart except for only one ULX, M101 X-1, for which mass of \mbh$=20-30$\msun\ was recently measured \citep{liu13}.
For this reason, the \mbh\ hosted by ULXs is still unknown and debated. 
ULXs were historically detected first in the X-rays and have often a wealth of observations in this energy band. 
Importantly, since the X-ray emission is produced and reprocessed in the innermost, hottest nuclear regions via Comptonization, the X-rays directly trace the activity of BHs and carry information from the inner core regions without being substantially affected by absorption.
Therefore, X-ray-based methods may provide an alternative way to constrain \mbh .
 
Recently, \citet{shapo09} developed a new method (hereafter, the X-ray scaling 
method) to determine \mbh\ and distance for GBHs using solely X-ray data. 
This method is based on self-similar trends found in the $\Gamma-{\rm QPO}$ and $\Gamma-N_{\rm BMC}$ diagrams (where $\Gamma$ is the X-ray photon index, QPO is the frequency of the quasi periodic oscillation, and $N_{\rm BMC}$ is the normalization of the bulk motion Comptonization model). 
The similarity shown by different GBHs in different outbursts allows one to determine the \mbh\ and distance by scaling the X-ray properties of a reference GBH, whose distance and $M_{\rm BH}$ are well constrained.
In our recent work, we demonstrated that the scaling method can be extended to larger scales to measure the \mbh\ in AGNs. To this end we used a sample of AGNs whose \mbh\ was already well constrained thanks to the reverberation mapping (RM) technique. 
The \mbh\ values determined using the \gam$-$\norm\ diagram are within a factor of 2-3 from the RM values suggesting that the scaling method is a reliable and robust \mbh\ estimator for any BH system accreting at moderate/high level \citep{glio11}. 
More specifically, the method is valid for AGNs with the accretion rate above 1\% in Eddington units and $L_{\rm X}$ above $10^{43}$ erg s$^{-1}$ \citep{jang2014}.

In this paper, we will apply the X-ray scaling method to estimate the \mbh\ for a sample of ULXs that possess multiple good-quality X-ray observations that are necessary to build \gam$-$\norm\ diagrams. 
The \mbh\ values determined in this manner will be compared with the values obtained with alternative methods described in the literature. 
For example, one of the most reliable estimates of the $M_{\rm BH}$ in ULXs is based on the detection of QPOs \citep[e.g.,][]{kaa09, RFK2010}.
However, this method is restricted to only a handful of ULXs for which QPOs have been robustly detected \citep{FR11}. 
The \mbh\ in ULXs is also measured from the relationship between the luminosity of the source and the inner disk temperature when the X-ray spectrum is parameterized by a multi-colored disk (MCD) model assuming that ULXs have the standard accretion disk and it extends to the last stable orbit \citep{FK09}. 
However, different spectral studies reveal that not all ULXs follow the expected $L_{\rm disk}\propto T^4$ relationship, suggesting that in some ULXs the X-ray emission cannot by described by a standard accretion disk.

This paper is structured as follows. In Section 2, we describe the characteristics of the ULX sample and report the data reduction procedure. The description of the X-ray scaling method and the results are given in Section 3. Section 4 contains the discussion of the main finding and our conclusions.

\section{Sample Collection \& Data Reduction}
\subsection{Sample collection}
We searched the \xmm\ archive for all ULXs from the Walton et al. 2011 catalog that were observed with at least 10 ks exposures and had been observed at least two number of times by \xmm .
A few more sources where added later on searching the \xmm\ archive. 
Our sample comprises 49 ULXs located in 22 nearby galaxies.
In Table \ref{tab1} column (1) indicates the host galaxy name, column (2) the ULX name, columns (3) and (4) the equatorial coordinates Right Ascension and declination, and column (5) the distance in units of Mpc. 
We used the distance value from the literature if available or the average distance of the host galaxy obtained from NED\footnote{http://ned.ipac.caltech.edu}. 

\begin{table}
\footnotesize
\begin{adjustwidth}{-0.7cm}{}
\begin{center}
\caption{The ULX Sample}
\begin{tabular}{llcccc} 
\hline        
\hline
\noalign{\smallskip}
\multicolumn{1}{c}{Host galaxy}&\multicolumn{1}{c}{ULX}&RA& Dec&$d$ (Mpc)\\
\multicolumn{1}{c}{(1)}&(2)&(3)&(4)&(5)\\
\hline
\noalign{\smallskip}
NGC 55 & ULX & 00:15:28.9 & -39:13:19.1 & 1.94\\
M31 & X-1 & 00:42:22.9 & 41:15:35.1 & 0.82\\
NGC 253 &X-1 & 00:47:22.6 & -25:20:51.0 & 3.19\\
& X-2 & 00:47:33.0 & -25:17:50.0 &\\
& XMM4 & 00:47:23.3 & -25:19:06.5\\
& XMM5 & 00:47:17.6 & -25:18:21.1 & \\
NGC 300 & XMM1 & 00:55:09.9 & -37:42:13.9 & 1.98\\
&XMM2 & 00:55:10.6 & -37:48:36.7 &\\
&XMM3 & 00:54:49.7 & -37:38:53.8 &\\
M33 & X-8 & 01:33:50.9 & 30:39:37.2 & 0.89\\
NGC 1313 & X-1 & 03:18:20.0 & -66:29:11.0 & 4.03\\
&X-2 & 03:18:22.3 & -66:36:03.8 &\\
&XMM2 & 03:17:38.8 & -66:33:05.3 &\\
&XMM4 & 03:18:18.5 & -66:30:05.0 &\\
IC 342 & X-1 & 03:45:55.5 & 68:04:54.2 & 3.12\\
& XMM2 & 03:46:15.0 & 68:11:11.2 &\\
& XMM3 & 03:46:48.6 & 68:05:43.2 &\\
& XMM4 & 03:46:57.2 & 68:06:20.2 &\\
NGC 2403 & X-1 & 07:36:25.9 & 65:35:38.9 & 3.54\\
HoII & X-1 & 08:19:29.0 & 70:42:19.0 & 3.33\\
M81 & X-6$^{\dagger}$ & 09:55:32.9 & 69:00:34.8 & 3.68\\
M82 & X-1 & 09:55:50.2 & 69:40:46.7 & 3.92\\
HoIX & X-1$^{\ddagger}$ & 09:57:53.2 & 69:03:48.3 & 3.63\\
NGC 4395 & XMM1 & 12:26:01.5 & 33:31:29.0 & 4.12\\
&XMM2 & 12:25:25.3 & 33:36:46.4 &\\
&XMM3 & 12:25:32.6 & 33:25:27.9 &\\
NGC 4490 & XMM1$^a$ & 12:30:32.4 & 41:39:14.6 & 8.68\\
&XMM2$^b$ & 12:30:36.5 & 41:38:33.3 &\\
&XMM3$^c$ & 12:30:43.3 & 41:38:11.5 &\\
&XMM4$^d$ & 12:30:31.1 & 41:39:08.1 &\\
&XMM5 & 12:30:30.3 & 41:41:40.3 &\\
NGC 4736 & XMM1 & 12:50:50.2 & 41:07:12.0 & 4.86\\
NGC 4945 & XMM1 & 13:05:33.3 & -49:27:36.3 & 3.98\\
& XMM2 & 13:05:38.4 & -49:25:45.3 &\\
& XMM3 & 13:05:18.8 & -49:28:24.0 &\\
& XMM4 & 13:05:22.2 & -49:28:27.9 & \\
& XMM5 & 13:05:25.7 & -49:28:32.3 & \\
NGC 5194 & XMM1 & 13:29:40.0 & 47:11:36.2 & 8.73\\
& XMM2 & 13:30:07.7 & 47:11:04.8 &\\
& XMM3 & 13:30:01.1 & 47:13:41.4 &\\
& XMM4 & 13:30:06.0 & 47:15:38.9 &\\
& XMM5 & 13:29:59.6 & 47:15:54.0 & &\\
& XMM6 & 13:29:57.5 & 47:10:45.3 &\\
& XMM7 & 13:29:53.6 & 47:14:31.5 &\\
NGC 5204 & X-1 & 13:29:38.6 & 58:25:06.0 & 5.28\\
NGC 5408 & X-1 & 14:03:19.6 & -41:23:00.0 & 4.85\\
NGC 5907 & ULX & 15:15:58.6 & 56:18:10.0 & 14.57\\
M101 & X-1 & 14:03:32.3 & 54:21:03.0 & 6.70\\
NGC 6946 & X-6 & 20:35:00.7 & 60:11:31.0 & 6.93\\
\hline
\end{tabular}
\label{tab1}
\end{center}
\footnotesize
{\it Note.} 
$^{\dagger}$M82 X-1 in \cite{hui2008}\\
$^{\ddagger}$M81 X-9 in \cite{tsunoda2006}\\
$^{a-d}$in the order of NGC 4490 X-4, X-6, X-8, and X-3 in \cite{yoshida2010} 
\end{adjustwidth}
\end{table}

\subsection{Data reduction}
We performed the data reduction following the standard procedures of Science Analysis System ({\small SAS}) version 12.0.1.
We only selected good X-ray event (``{\small FLAG}$=0"$) with patterns of $0-4$ and $0-12$ for the pn and MOS, respectively. 
Most of the ULXs in our sample were isolated point-like sources, whose emission can be clearly separated from the galactic nucleus contribution. 
For those targets, we used source extraction regions with a radius of $10"-20"$ and background regions of $\sim60"$ located in a nearby source free zone. 
Some observations captured the source either at the edge of CCD or partially in the gap between CCDs for the pn and/or MOS. 
In this case, the source extraction region was reduced accordingly.
When the source in the \xmm\ image did not appear to be isolated (e.g., when the ULX emission could be contaminated by diffuse emission or by nearby sources), we used {\it Chandra} images to guide our source extraction and assess the possible contamination. 
In general, the spectral analysis was performed by simultaneously fitting the spectra from the three EPIC cameras. 
Only for very bright sources the analysis was limited to the EPIC pn data.
The {\small RMFGEN} and {\small ARGEN} tasks were used to generate {\small RMF} and {\small ARF} files, respectively. 
To use the $\chi^2$ statistics, each spectrum was grouped with 20 counts per bin or 15 counts per bin in case of relatively short observations (net exposure $\sim10$ ks).

\section{\mbh\ Measurements}
\subsection{The X-ray scaling method}
\subsubsection{GBH reference outbursts}
\begin{figure*}
\footnotesize
\includegraphics[scale=0.5]{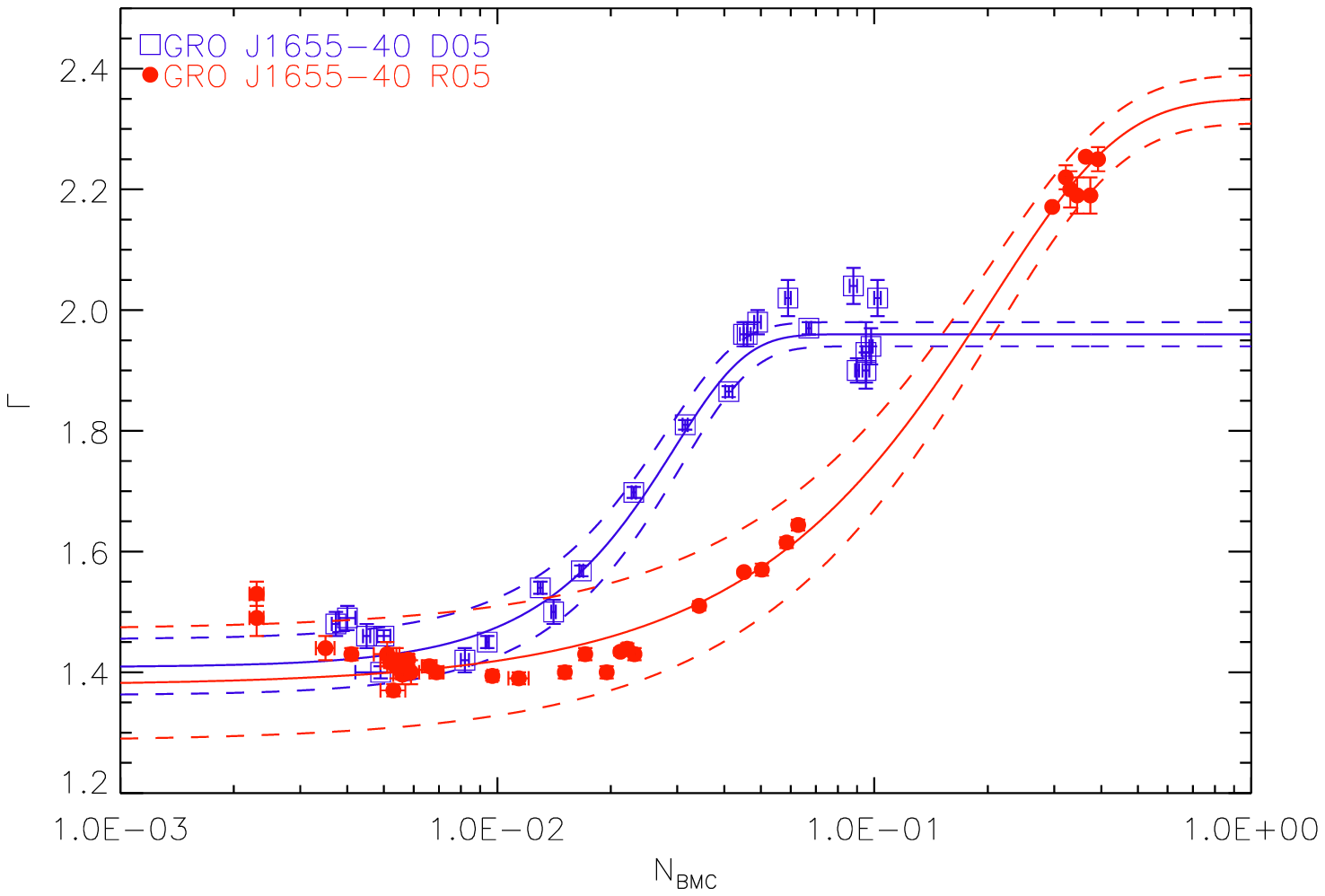}
\includegraphics[scale=0.5]{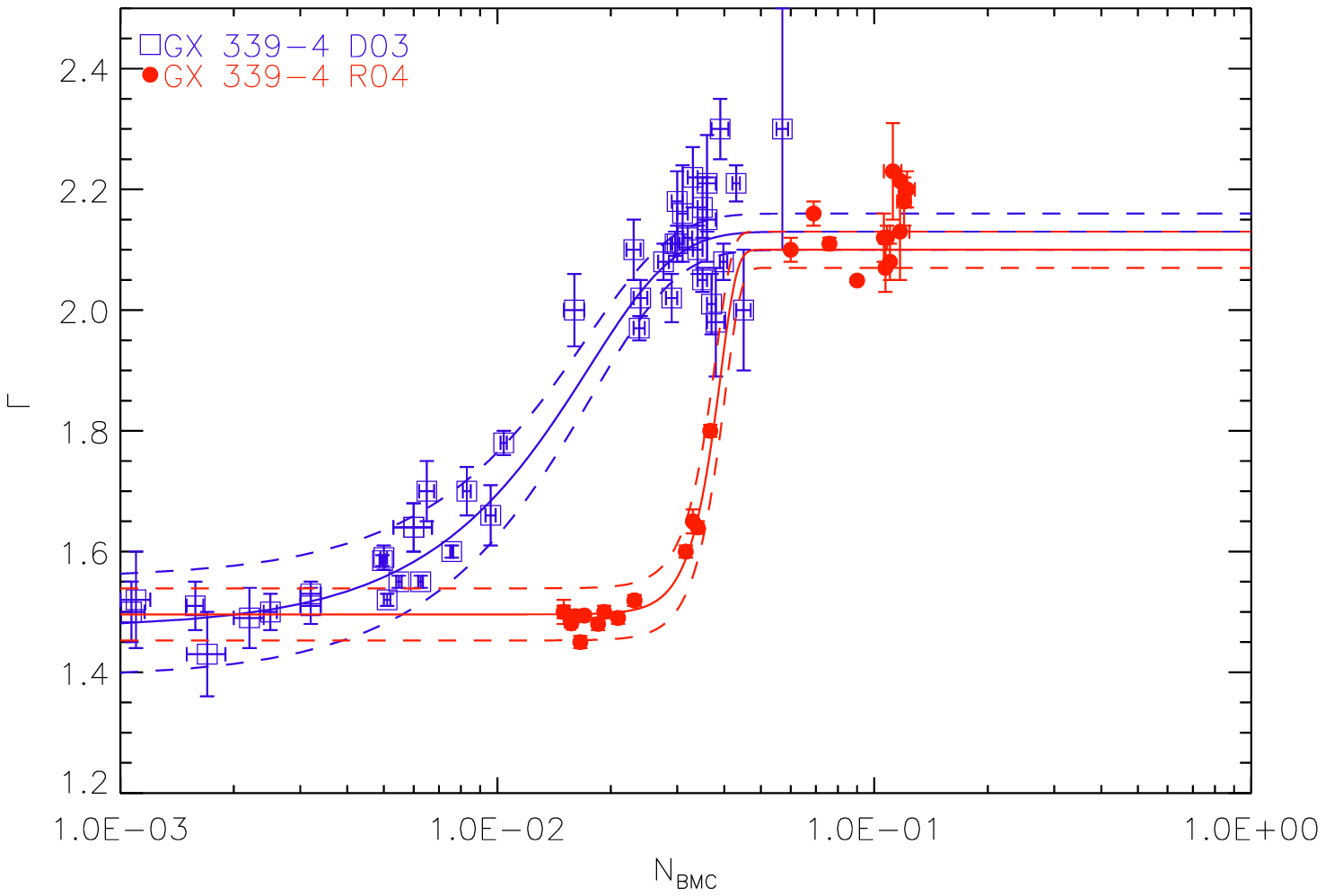}
\includegraphics[scale=0.5]{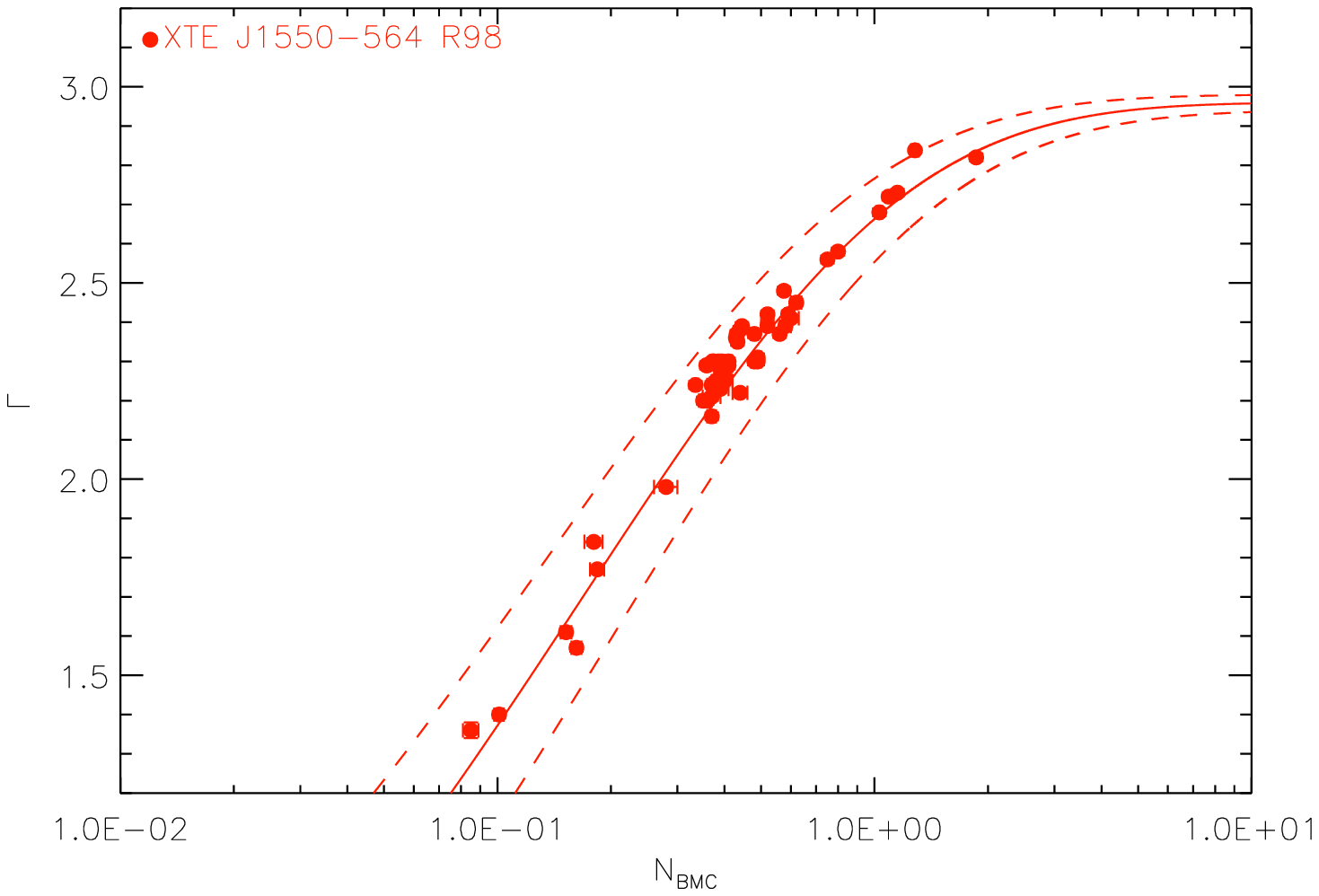}
\includegraphics[scale=0.5]{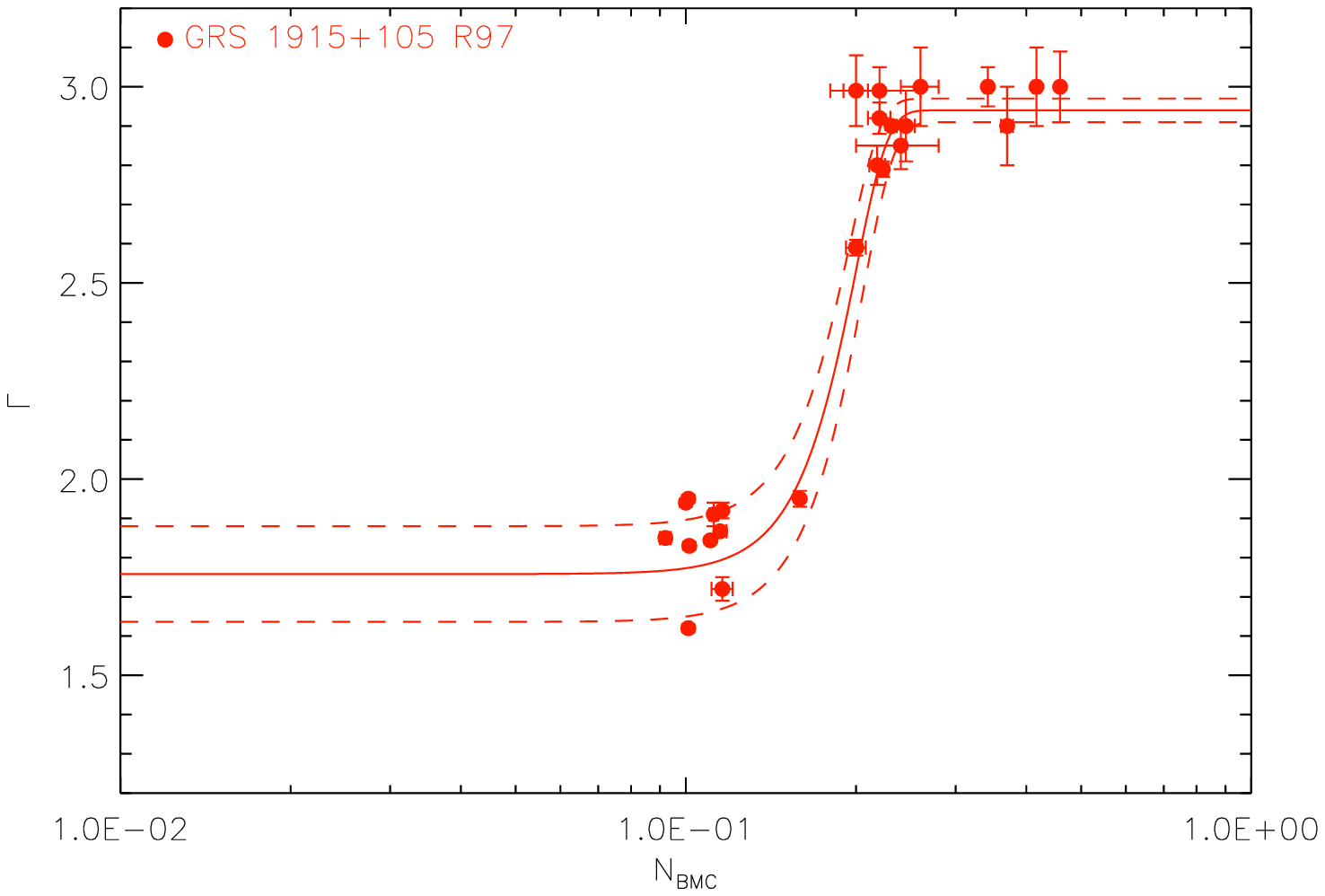}
\footnotesize
\caption[\gam$-$\norm\ diagrams of reference patterns]
{\footnotesize \gam$-$\norm\ diagrams of GBH reference sources.  
In each plot, the decay phase data points are indicated with the open squares (blue in color the version) and the filled circles (red) for the rise phases. 
The solid line indicates the best-fit using Eq. \ref{equ1} and the dashed lines its uncertainty within $1\sigma$.}
\label{referencepattern}
\end{figure*}

To determine the \mbh\ of a given target with the X-ray scaling method, its $\Gamma-N_{\rm BMC}$ diagram should be compared with analogous diagrams of GBH reference sources. To construct the latters we have used the spectral transitions of four different GBHs: two moderate accreting GBHs $-$ the decay (a transition from moderately accreting the high/soft to the low/hard states) and rise (from the low/hard to the high/soft states) phase transitions of GRO J1655$-$40 in 2005 (hereafter, GROJ1655D05 and GROJ1655R05, respectively) and the decay phase in 2003 and the rise phase in 2004 of GX 339$-$4 (GX339D03 and GX339R04, respectively) $-$ and two highly accreting GBHs $-$ XTE J1550$-$564 rise phase in 1998 (XTEJ1550R98) and GRS 1915$+$105 rise phase in 1997 (GRS1915R97). 
We present the $\Gamma-N_{\rm BMC}$ diagrams of reference patterns in Figure \ref{referencepattern}. 

Each spectral transition was fitted using the {\small Levenberg-Marquart} algorithm \citep{press1997} with the following function 
\begin{equation}
N_{\rm BMC}(\Gamma)=N_{\rm tr}\times\left[\ln\left
(\exp\left(\frac{A-\Gamma}{B}\right)-1\right)+1\right]^{1/\beta}
\label{equ1}
\end{equation}
where parameter $A$ characterizes the upper saturation level, $B$ the lower saturation level of the spectral evolution, $\beta$ the slope of pattern, and $N_{\rm tr}$ describes the shift of the function along the $x$-axis. 
Eq. \ref{equ1} is the inverse Eq. 10 of \gam(\norm) in \cite{shapo09}. 
The physical properties of reference sources, \mbh\ and the distance, and parameters of the fitting function are reported in Table \ref{tab2}. 

\begin{table*}
\footnotesize
\caption{Information of the reference patterns}
\begin{center}
\begin{tabular}{lcccccc} 
\hline        
\hline
\noalign{\smallskip}
\multicolumn{1}{c}{Reference pattern} & \mbh/\msun\ & $d$ (kpc) & $A$ & $B$ & $N_{\rm tr}$ & $\beta$ \\
\hline
\noalign{\smallskip}
GROJ16550D05 & $6.3\pm0.4$ & $3.2\pm0.2$ & $1.96\pm0.02$ & $0.42\pm0.02$ & $0.023\pm0.001$ & $1.8\pm0.2$\\
GROJ1655R05 & & &$2.35\pm0.04$ & $0.74\pm0.04$ & $0.131\pm0.001$ & $1.0\pm0.1$\\
GX339D03 & $12.3\pm1.4$ & $5.7\pm0.8$ & $2.13\pm0.03$ & $0.50\pm0.04$ & $0.013\pm0.001$ & $1.5\pm0.3$\\
GX339R04 & & & $2.10\pm0.03$ & $0.46\pm0.01$ & $0.037\pm0.001$ & $8.0\pm1.5$\\
XTE1550R98 & $10.7\pm1.5$ & $3.3\pm0.5$ & $2.96\pm0.02$ & $2.80\pm0.20$ & $0.055\pm0.010$ & $0.4\pm0.1$\\
GRS1915R97 & $12.9\pm2.4$ & $9.2\pm0.2$ & $2.94\pm0.03$ & $0.9\pm0.07$ & $0.186\pm0.005$ & $6.1\pm1.9$\\
\hline
\hline
\end{tabular}
\end{center}
\footnotesize
\label{tab2}
\begin{itemize}
\item[] {\it Notes.} $-$ $A$, $B$, $N_{\rm tr}$, and $\beta$ are the parameters of the 
function, described Eq. 1, used to fit the spectral patterns in the 
$\Gamma-N_{\rm BMC}$ diagram.
\end{itemize}
\end{table*}

Each reference pattern carries its own advantage in this study. 
GRO J1655-40 is the best-parameterized system so the scaling method will reduce its uncertainty in the \mbh\ estimation. 
GX 339-4 is the prototypical GBH having very similar spectral variability from different outbursts. 
XTE J1550-564 has the largest photon index range (\gam\ $=1.3-3$) and GRS1915R97 whose has a high $\Gamma$ saturation level ($\Gamma=3$) which can be used as an additional reference pattern for the targets with high $\Gamma$. 

\subsubsection{ULX application}
Since QPO features in ULXs are elusive due to the low signal-to-noise ratio of the X-ray data available and have been detected from only a few ULXs, the \mbh\ determination using the $\Gamma-{\rm QPO}$ diagram is fairly limited. 
However, we can use the $\Gamma-N_{\rm BMC}$ diagram as we did for AGNs (Gliozzi et al. 2011). 
The shorter time scales expected in ULXs (in comparison to those associated with AGNs) have the potential to probe the spectral evolution of ULXs over time intervals of months/years and to allow a direct comparison with GBH reference outbursts. 
If the spectral evolution in any ULXs is similar to GBHs in \gam$-$\norm\ diagram, then \mbh\ can be characterized by the horizontal shift of the trend of GBH until it reaches the trend of the ULX. 
This horizontal shift is directly associated to the change of the parameter $N_{\rm tr}$ (see Eq. \ref{equ1}) and can be measured from the best-fit of spectral transitions of unknown ULXs after fixing the parameter $A$, $B$, and $\beta$ to the best-fit values of the reference sources. 

\begin{figure}
\includegraphics[scale=0.5]{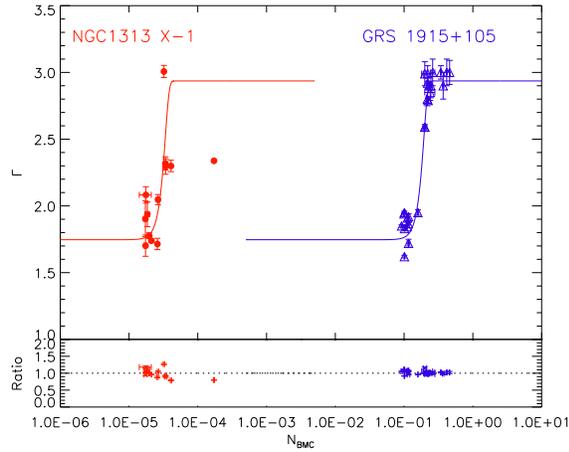}
\footnotesize
\caption[A plot of \gam$-$\norm\ for GRS 1915+105 and NGC 1313 X-1]{\footnotesize A plot of \gam$-$\norm\ for GRS 1915+105 and NGC 1313 X-1. The spectral properties of NGC 1313 X-1 are plotted with filled circles (red in color) and open triangles (blue) for GRS 1915+105. In the bottom panel, we plotted the ratio between the data points and the best-fit versus \norm.}
\label{fig1}
\end{figure}

Figure \ref{fig1} illustrates the scaling method with the ULX NGC 1313 X-1 and GRS 1915$+$105 as reference source. 
The spectral evolution of NGC 1313 X$-$1 with observations from 2000 to 2006 is plotted and is described by best-fitting pattern of GRS1915R97 with all parameters fixed except for $N_{\rm tr}$. 
The best-fit was done using the {\small IDL} software package called {\small LMFIT} which accounts for errors on the both axes. 
The good-fit of NGC 1313 X$-$1 was also visually confirmed by the plot of the ratio between the best-fit and the data point versus \norm\ which is shown \gam$-$\norm. 
The basic steps of the X-ray scaling method applied to ULXs can be summarized as follows.
\begin{itemize}
\item[(1)] Systematically fit all energy spectra of the ULXs with the BMC model.
\item[(2)] Construct the \gam$-$\norm\ diagrams for ULXs of known 
distance and compare them to all the GBH reference patterns. 
\item[(3)] Measure the best-fit $N_{\rm tr}$ value for ULX using 
Eq. \ref{equ1} and determine the \mbh\ value from the equation below
\begin{equation}
\,\,\,\,\,\,\,\,\,\,\,\,\,\,\,\,\,\,\,\,\,\,\,\,\,\,\,\,\,\,
M_{\rm BH,t}=M_{\rm BH,r}\times\frac{N_{\rm tr,t}}
{N_{\rm tr,r}}\times\left(\frac{d_{\rm t}}{d_{\rm r}}\right)^2
\label{equ2}
\end{equation}
where subscribed t and r are used for the target and reference source, 
and $d_{\rm t}$ and $d_{\rm r}$ are the respective distances. 
\end{itemize}

\subsection{Spectral analysis}
Each \xmm\ spectrum in the range of $0.5-10$ keV was systematically fitted using the X-ray astronomy 
software package {\small XSPEC} V12.0.1. 
Two absorption models were used fixing one model at the Galactic value and setting the other one as a free variable to mimic the intrinsic local absorption. 
The soft X-rays were described by an accretion disk model called {\small diskpn} \citep{gierlinski1999} that is parameterized by the maximum disk temperature near the black hole $T_{\rm max}$ in the units of keV and the inner disk radius in the units of $R_g$. 
To fit the hard X-rays, which are thought to be produced via Comptonization, we used the BMC model, whose parameters are: 
the temperature of seed photons, $kT$, the energy spectral index $\alpha$ (related to \gam\ by the relation of $\Gamma=1+\alpha$), $\log(A)$ which is related to the \comp\ fraction by $f=A/(1+A)$ (i.e., the ratio of the number of scatted photons to the seed photons), and the normalization \norm\ which is directly related to the luminosity and inversely to the square of the distance. 
We used Gaussian models when there were line-like features. 

We set the seed photon temperature equal to  $T_{\rm max}$, $\log(A)$ at 2 if its initial best-fit value was $\gg2$ and the inner disk at the last 
stable orbit with the minimum of $6R_g$. 
We used the F-test to check the significance of the different model components. 
The spectral result was considered acceptable when the reduced $\chi^2$ was in the $0.8-1.5$ range. 

\begin{figure*}
\footnotesize
\includegraphics[scale=0.5]{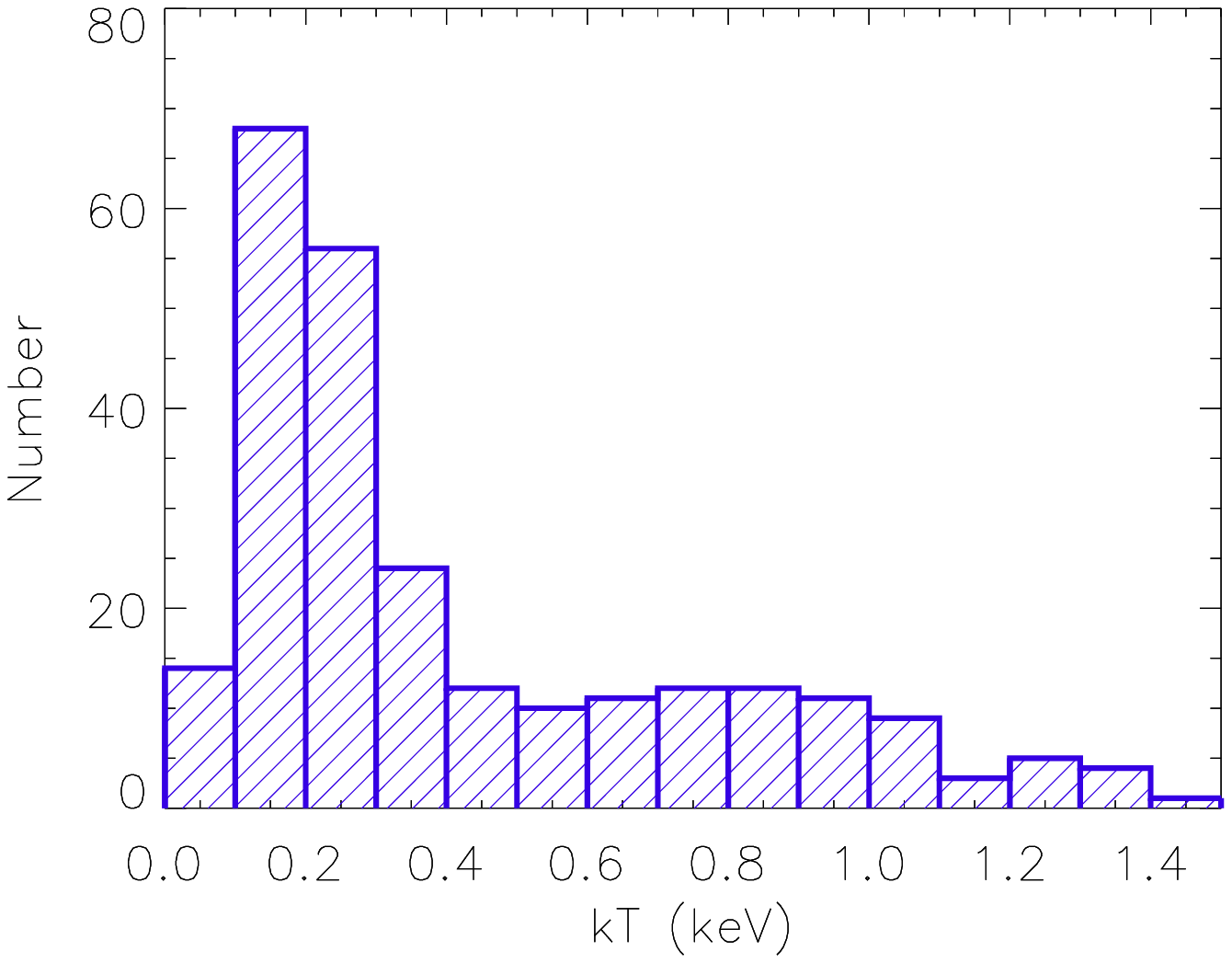}
\includegraphics[scale=0.5]{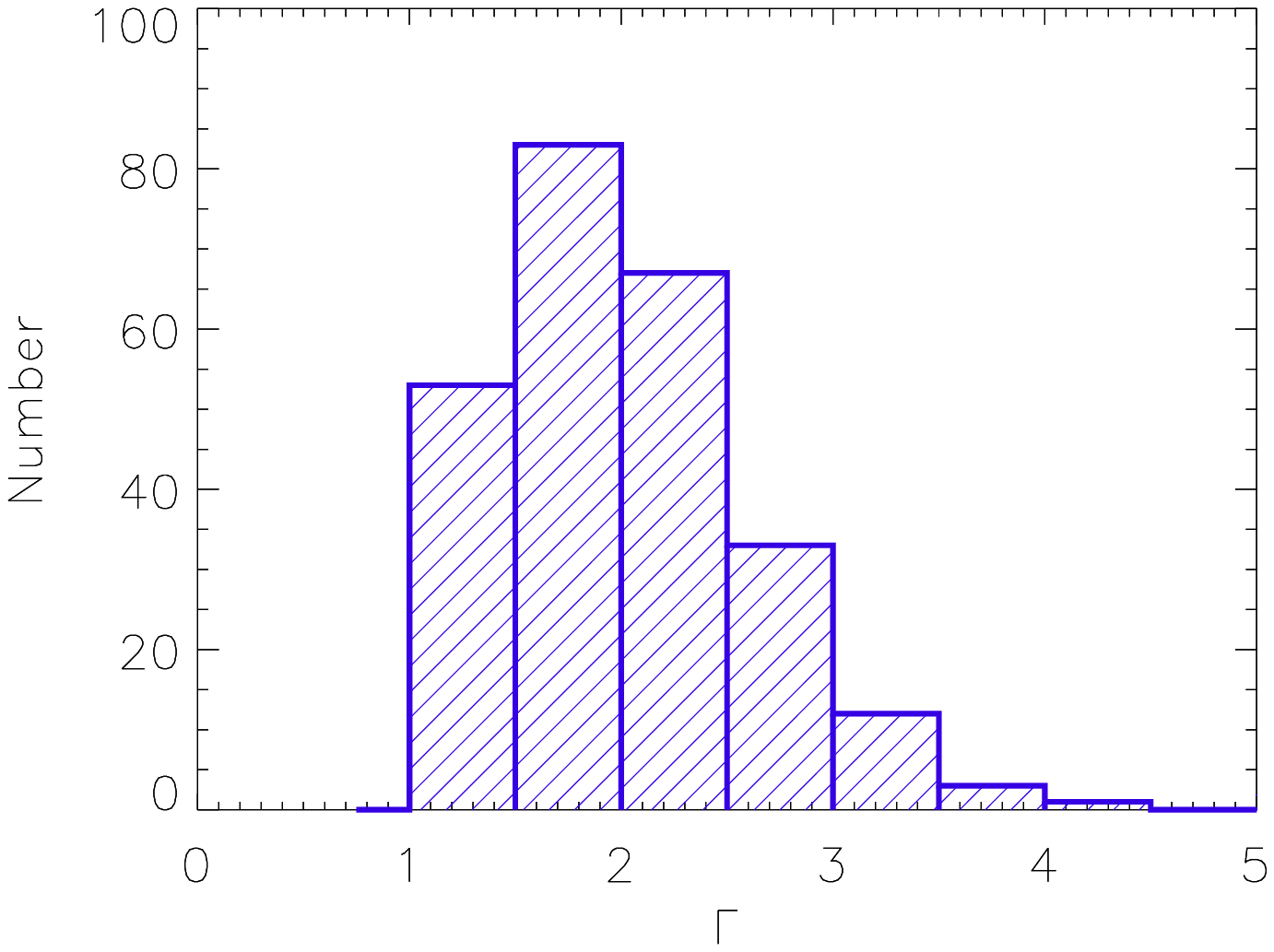}
\includegraphics[scale=0.5]{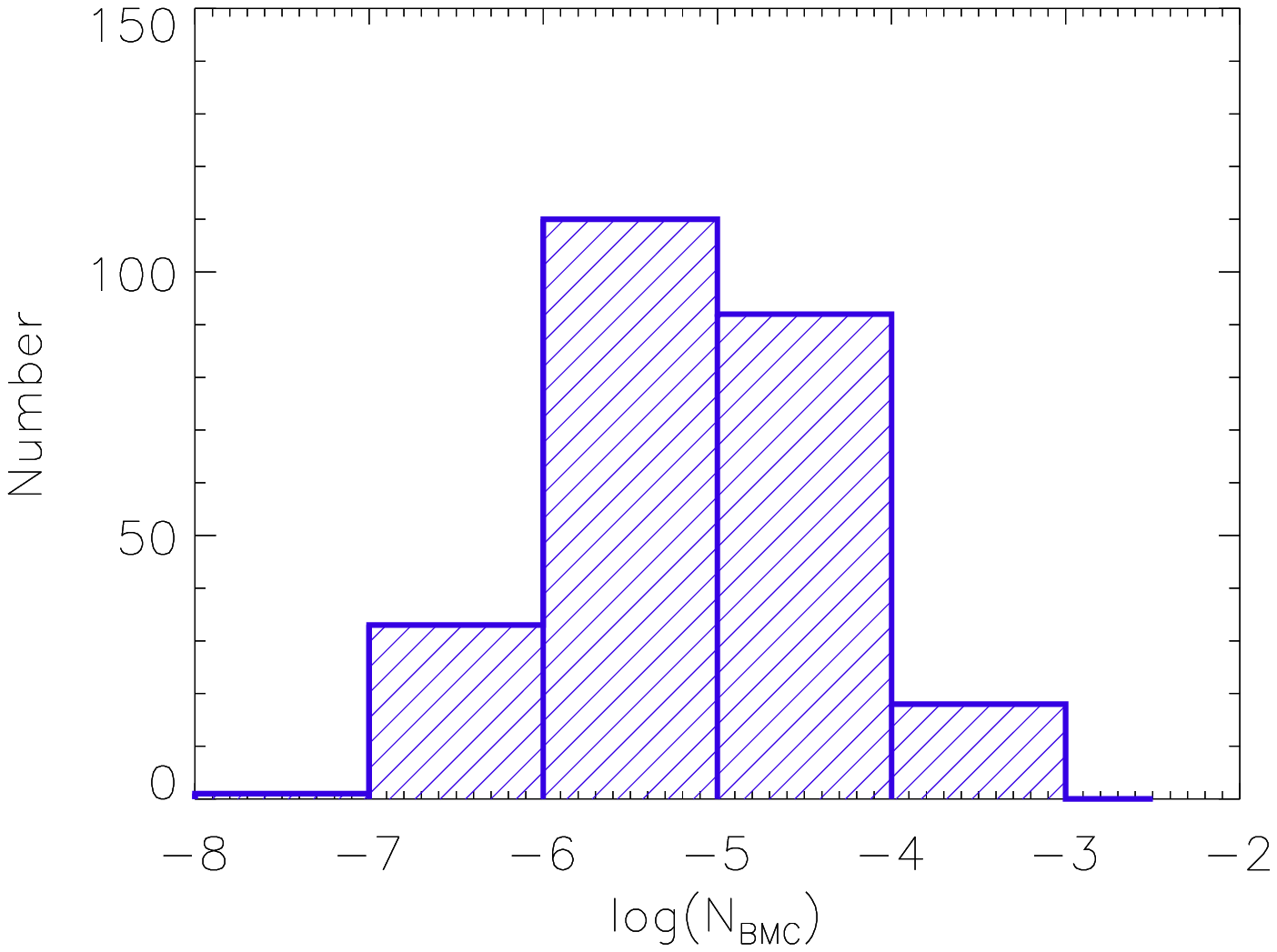}
\includegraphics[scale=0.5]{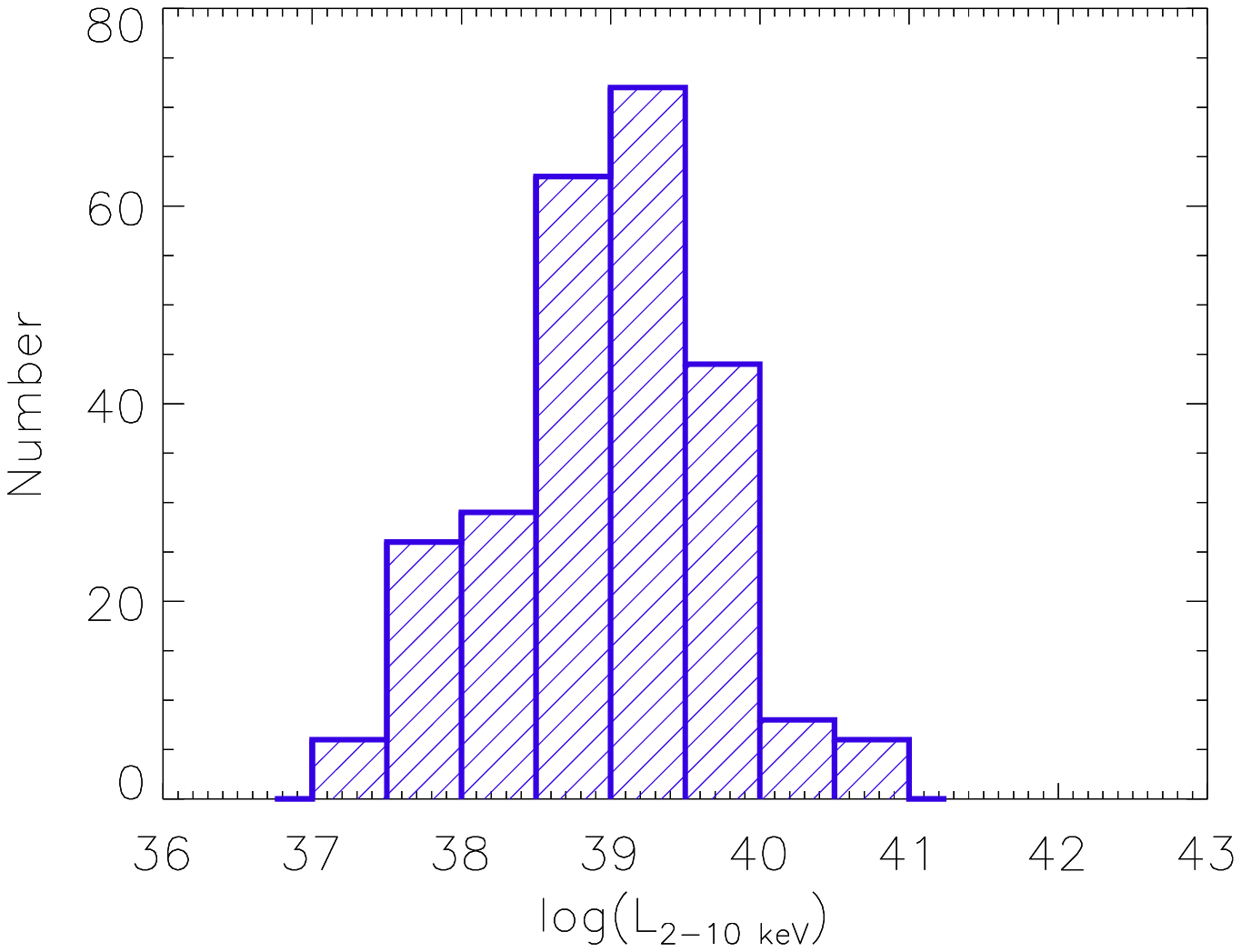}
\footnotesize
\caption[Histograms of spectral results]{Histograms of spectral results: 
distributions of $kT$, \gam, \norm, and $\log(L_{2-10\,\rm keV})$ for all 
260 observations from 47 ULXs.}
\label{fig2}
\end{figure*}

The spectral results for 48 ULXs (with a total of 262 observations) are illustrated in Figure \ref{fig2}, where the distribution of $kT$, $\Gamma$, $\log(N_{\rm BMC})$, and $\log(L_{\rm X})$ are shown. 
The measured $kT$ ranges between $0.01-1.79$ keV with a mean of $0.43\pm0.35$ keV and a median of $0.26$ keV. 
The vast majority of the spectra (206 out of 262 observations) had $\log(A)$ fixed at 2 and a mean value of $1.64\pm0.71$. 
The value of $\Gamma$ are distributed in the $1-6$ range with the mean value of $2.04\pm0.68$ where 244 observations have $\Gamma<3$. 
The $N_{\rm BMC}$ is distributed in the $10^{-4}-10^{-8}$ range with a mean of $(2.44\pm4.61)\times10^{-5}$ and a median of $3.21\times10^{-6}$. 
The unabsorbed luminosity in $2-10$ keV was in the $10^{37}-10^{41}$ \ergs\ range.

\subsection{$\Gamma-N_{\rm BMC}$ diagrams}
We constructed $\Gamma-N_{\rm BMC}$ diagrams for each ULX to investigate their spectral evolution and constrain their \mbh. 
Out of 48 ULXs, 5 ULXs (NGC 55 ULX, NGC 253 XMM4, NGC 4490 XMM2, NGC 4945 XMM4, and XMM5) have values of \gam\ outside the range of any reference pattern. 
We did not construct the $\Gamma-N_{\rm BMC}$ diagram for NGC 4945 XMM3 because there was only one observation with good \xmm\ quality data. 
Therefore, these ULXs were excluded from further analysis since their \mbh\ cannot be constrained with this method. 

\begin{figure*}
\includegraphics[scale=0.5]{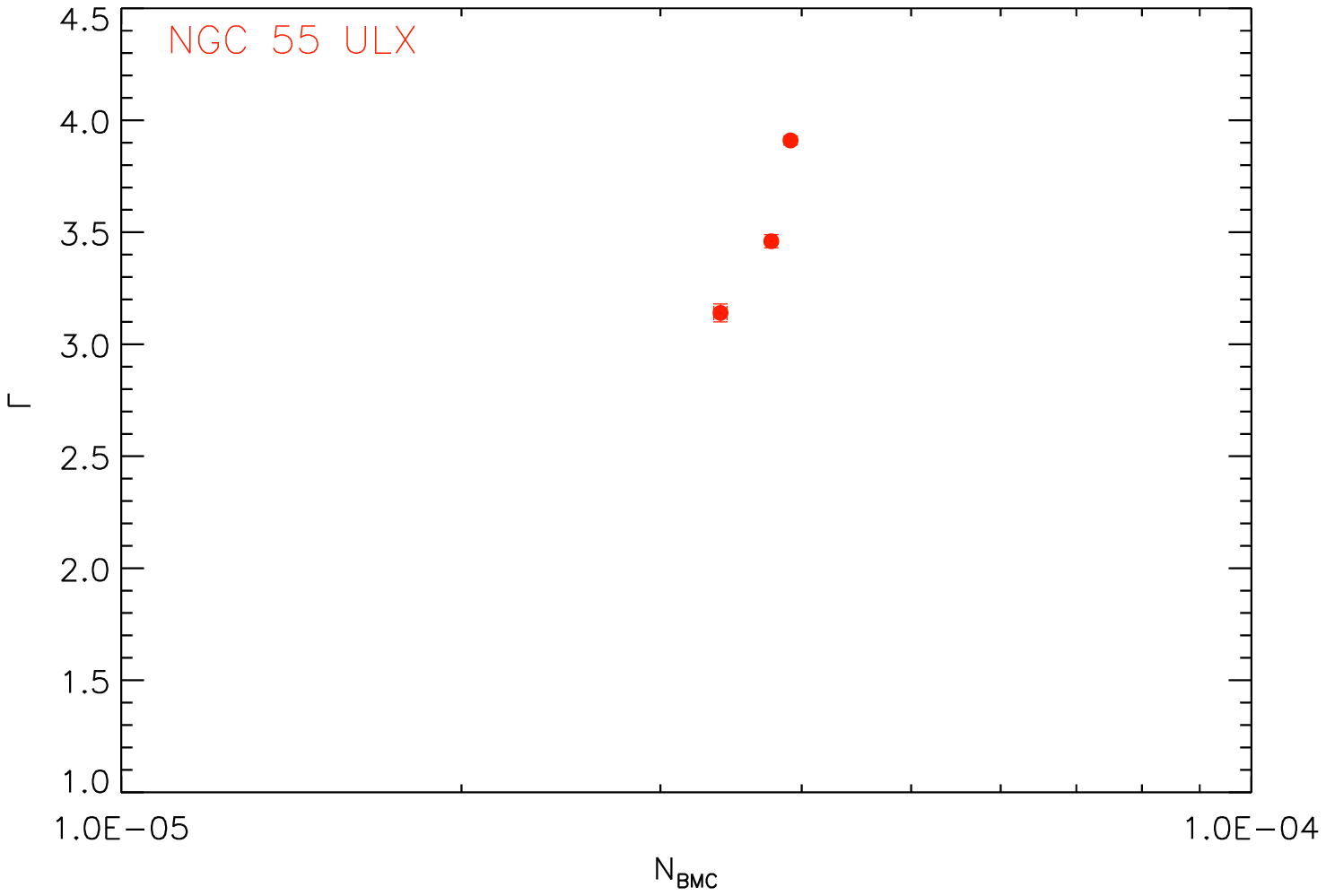}
\includegraphics[scale=0.5]{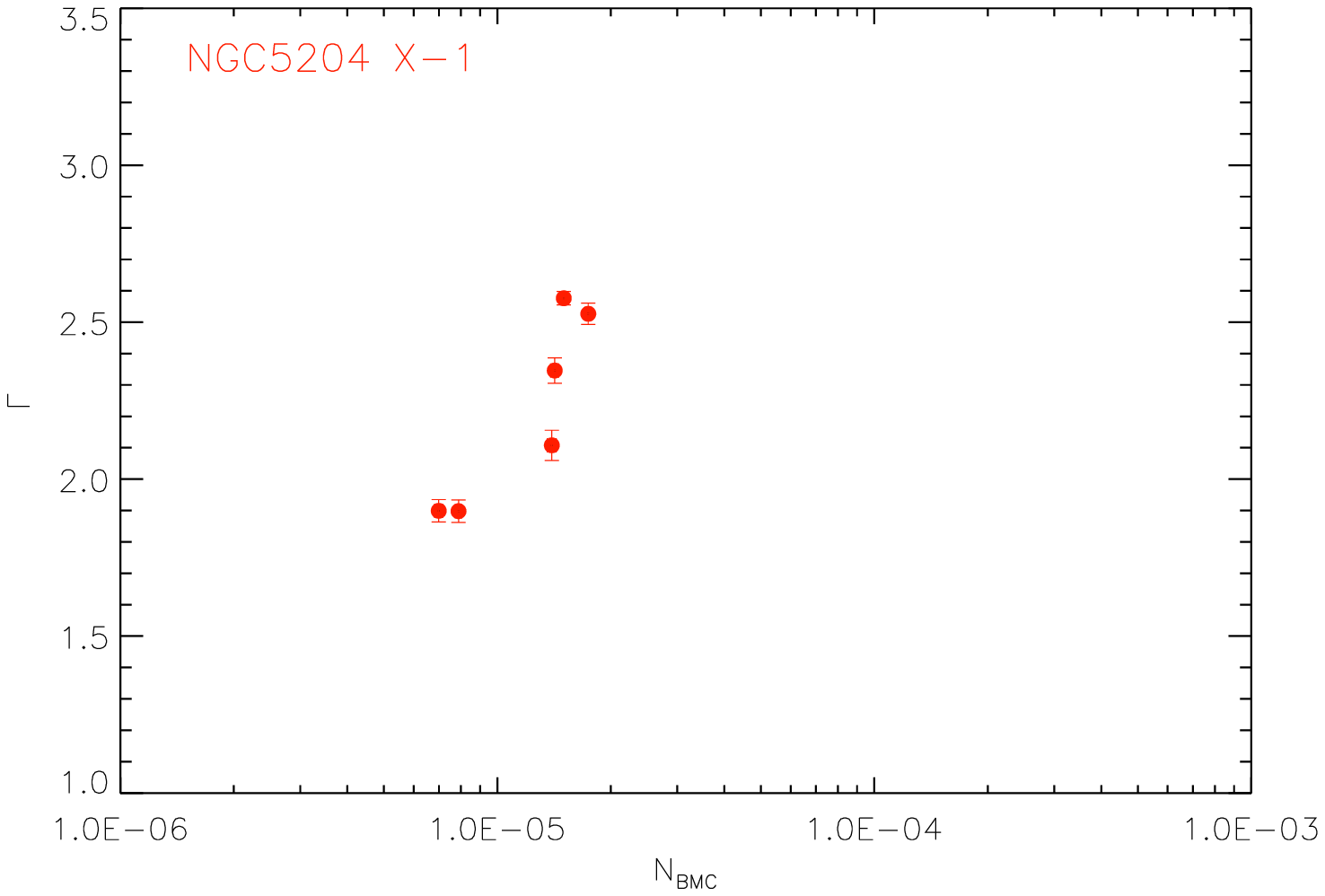}
\includegraphics[scale=0.5]{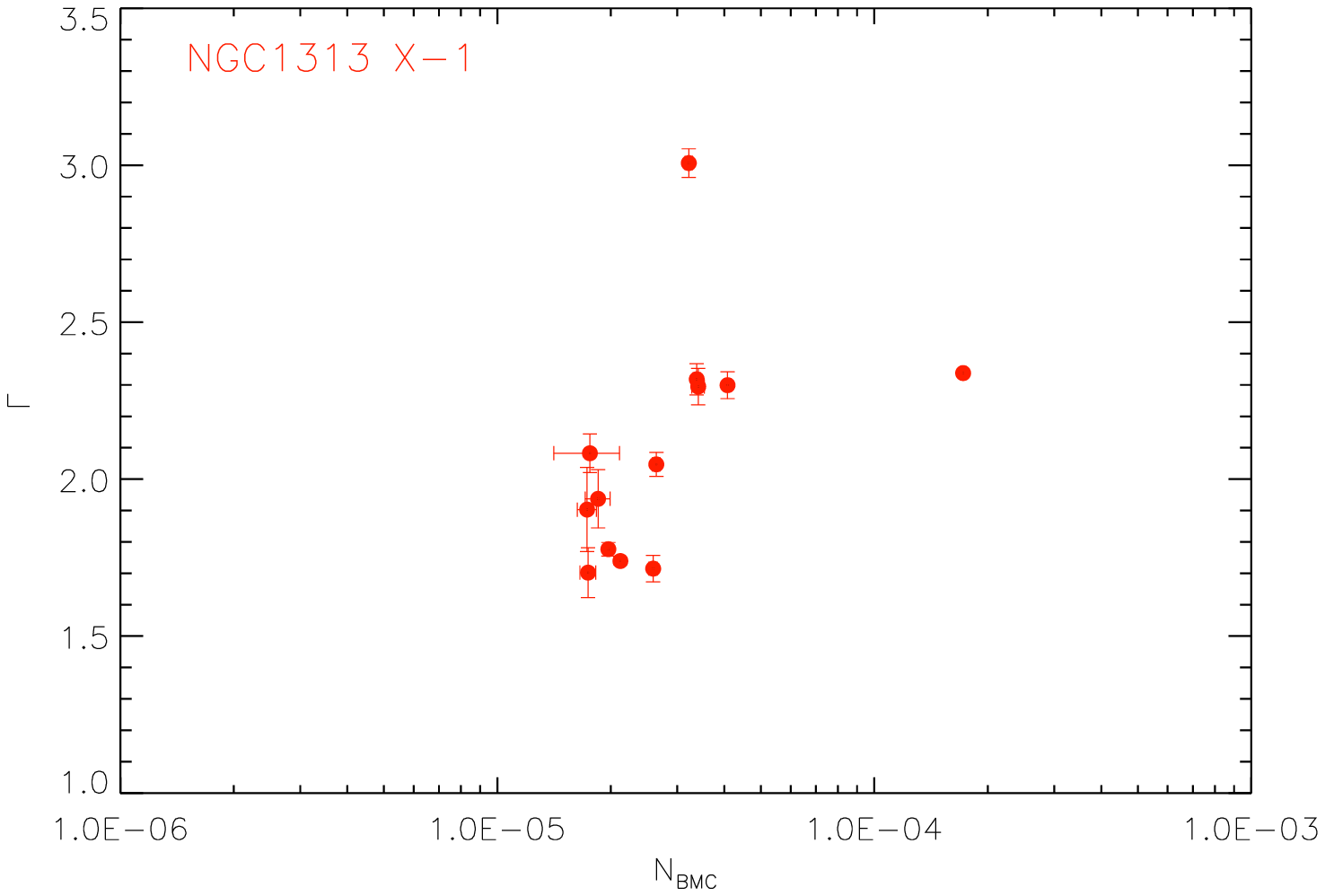}
\includegraphics[scale=0.5]{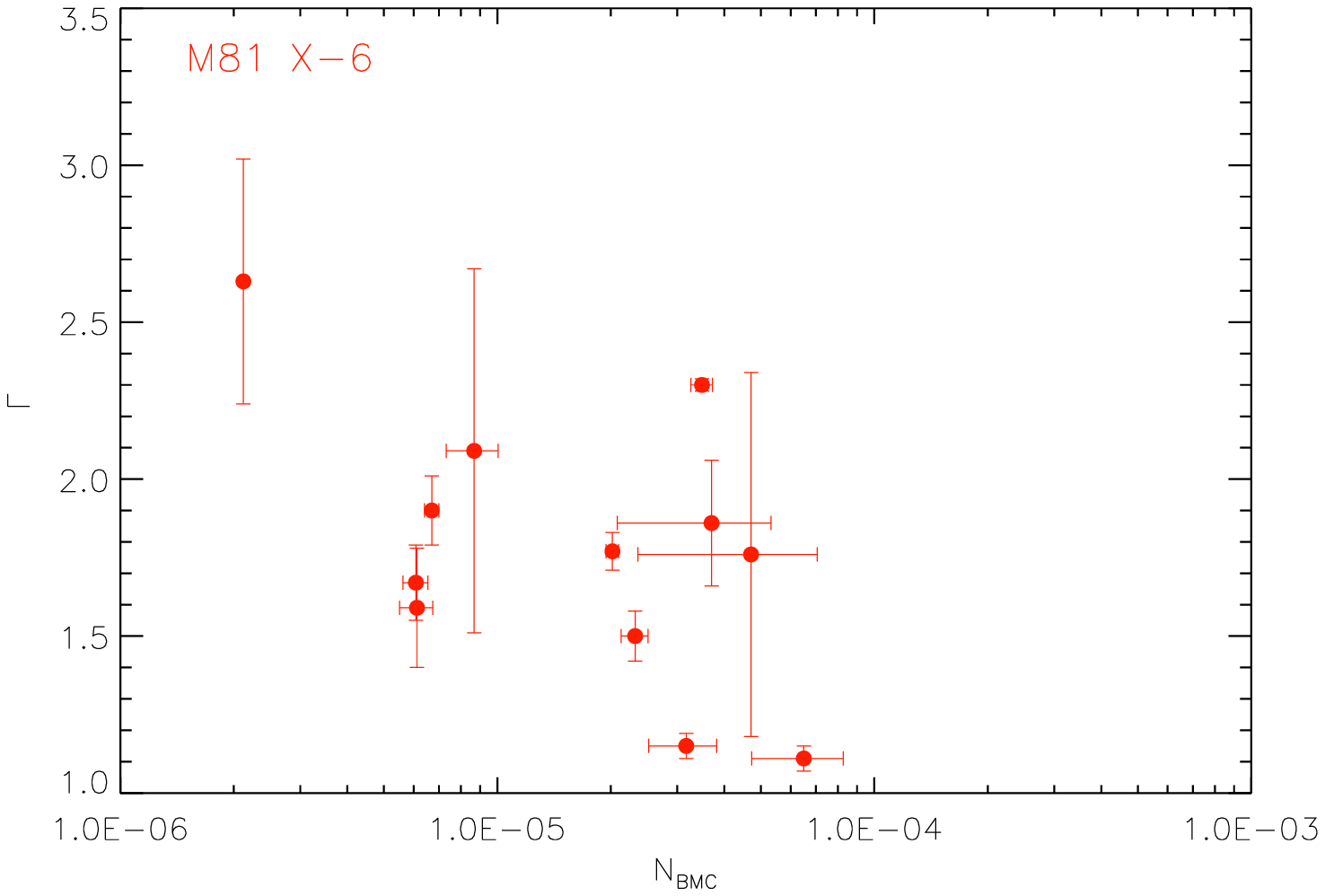}
\caption[$\Gamma-N_{\rm BMC}$ diagram of ULXs]
{\footnotesize $\Gamma-N_{\rm BMC}$ diagram of ULXs. 
We show different types of $\Gamma-N_{\rm BMC}$ patterns of NGC 55 ULX, NGC 5204 X-1, NGC 1313 X-1, and M81 X-6.}
\label{GNormExample}
\end{figure*}

28 ULXs showed positive $\Gamma-N_{\rm BMC}$ trends, whereas 15 ULXs had irregular or negative patterns. 
Some ULXs exhibit different values of $\Gamma$ corresponding to the same value of $N_{\rm BMC}$. 
Figure \ref{GNormExample} illustrates the different types of trends shown; 
NGC 55 ULX has all values of $\Gamma$ above 3 and therefore cannot be compared to any reference pattern. 
NGC 5204 X-1 shows a positive spectral pattern in the $\Gamma=1.7-2.5$ range and can be compared to any reference pattern. 
NGC 1313 X-1 shows a positive spectral trend with two possible outliers. 
Finally, an anti-correlation of $\Gamma$ and $N_{\rm BMC}$ appears to be present in $\Gamma-N_{\rm BMC}$ plot of M81 X-6. 

\subsection{\mbh\ computation}
The spectral trends of each ULX in the $\Gamma-N_{\rm BMC}$ plot were fitted with the same parametric function used to describe each reference trend. 
The parametric function (see Eq. \ref{equ1}) used for the fit  is characterized by three parameters ($A$, $B$, and $\beta$) fixed at the reference source values, and by the parameter $N_{tr}$ that describes the shift along the $x$-axis of $\Gamma-N_{\rm BMC}$ diagram is left free to vary. 
We tried to compare as many reference patterns as possible to each ULX trend: 13 ULXs were compared to all 6 reference patterns, 12 ULXs to 5, 3 ULXs to 4, 6 to 3, 6 to 2, and 4 ULXs to only one reference pattern. 

In the case where the ULX shows a clear positive correlation between \gam\ and $N_{\rm BMC}$ with a trend similar to one of the references, the computation of $M_{\rm BH,Scale}$ was straightforward. 
Irregular patterns can be explained by a combination of different events over several years  (keep in mind that also for the reference sources, the spectral pattern during the rise phase of the outburst may be different from the decay trend). 
Alternatively, irregular patterns can be explained by the presence of statistical outliers or by spurious points obtained from low signal-to-noise spectra. 
The $\Gamma-N_{\rm BMC}$ diagram of NGC 1313 X-1 (see Figure \ref{fig1} and \ref{GNormExample}) show an example of an ULX with two apparent outliers one at $\Gamma=3$ and at $N_{\rm BMC}\ge10^{-4}$. 

\begin{figure}
\includegraphics[scale=0.5]{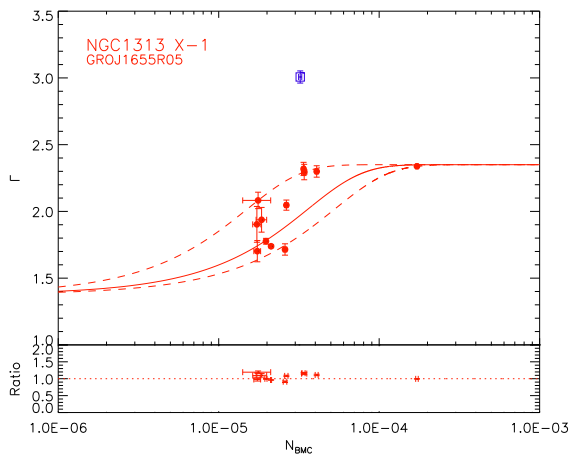}
\caption[NGC 1313 X-1 \gam$-N_{\rm BMC}$ diagram]
{\footnotesize The \gam$-N_{\rm BMC}$ diagram of NGC 1313 X-1 fitted with GROJ1655R05 pattern. We excluded the point at $\Gamma\approx3$ that is indicated with the open square (blue).}
\label{NGC1313-Example-02}
\end{figure}

When the spectral transition of NGC 1313 X-1 was fitted with the GRS1915R97 pattern, we included the point (at $\Gamma=3$) as a part of the pattern whereas the point with $N_{\rm BMC}>1\times10^{-4}$ was treated as an outlier and excluded from the $M_{\rm BH}$ computation (see Figure \ref{fig1}).  
The data point at $\Gamma=3$ was an outlier instead when its spectral transition was best-fitted using the reference GROJ1655R05 (see Figure \ref{NGC1313-Example-02}).

\begin{figure*}
\includegraphics[scale=0.5]{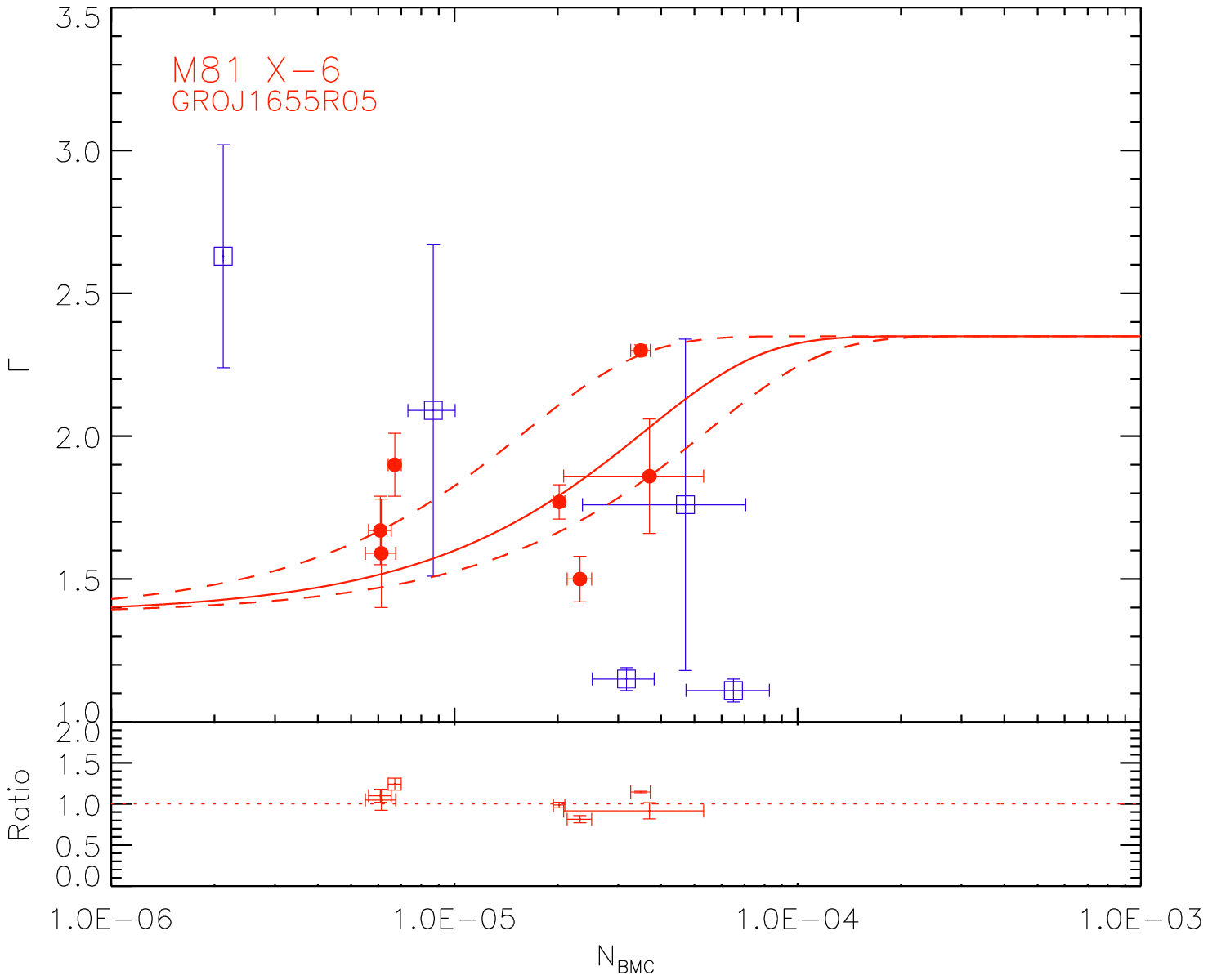}
\includegraphics[scale=0.5]{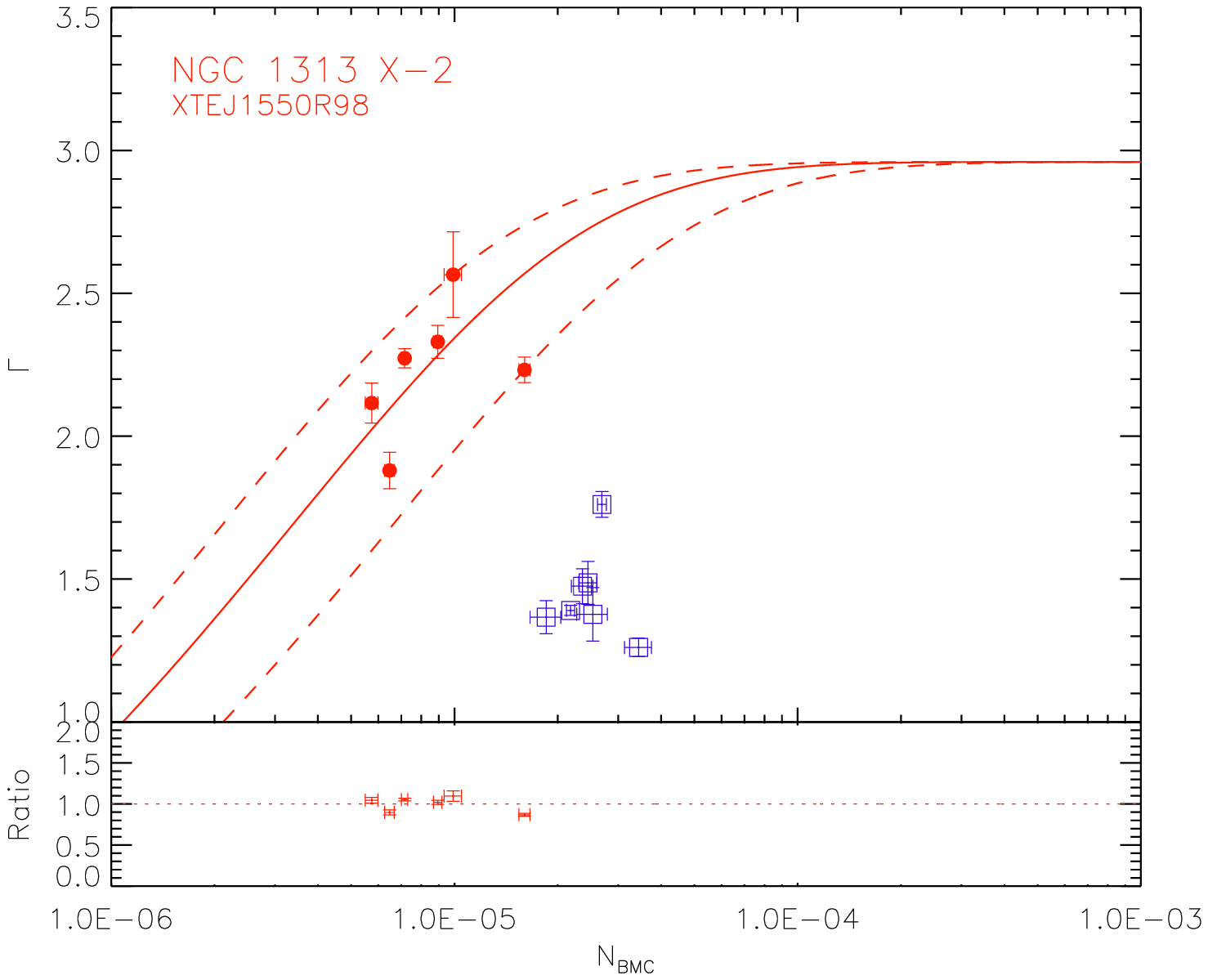}
\caption[\gam$-N_{\rm BMC}$ diagram]
{\footnotesize \gam$-N_{\rm BMC}$ diagram. 
The $\Gamma-N_{\rm BMC}$ diagram of M81 X-6 fitted by GROJ1655R05 pattern is in the left panel and NGC 1313 X-2 fitted by XTEJ1550R98 in the right panel. 
The data points used in the best-fitting are indicated with filled circles and the excluded ones with open squares. 
The ratio between points and the best-fit is also plotted in the bottom of $\Gamma-N_{\rm BMC}$ diagram.}
\label{GNormBest-FitExample02}
\end{figure*}

The left panel of Figure \ref{GNormBest-FitExample02} shows the apparently complex pattern of M81 X-6. 
However, after we exclude the data points with huge error-bars and those with unphysically low values of \gam, the remaining data can be fitted with the usual positive trend shown by the reference sources. 
The right panel plot of Figure \ref{GNormBest-FitExample02} shows the case of NGC 1313 X-2 which appears to have two separate clusters of data. 
Once again, excluding the data characterized by low values of \gam\ (which cannot be compared to any reference trend) makes it possible to fit the remaining data with one of the standard patterns shown by the reference sources. 

We computed the uncertainty of \mbh\ values based on the uncertainty of parameters from the reference patterns. We accepted the $1\sigma$ uncertainty values of the best-fit when all data points were within the range. Otherwise, we expanded the boundaries of the best-fit to visually confirm that all data points are in the range of $2-3\, \sigma$.

\begin{figure*}
\includegraphics[scale=0.5]{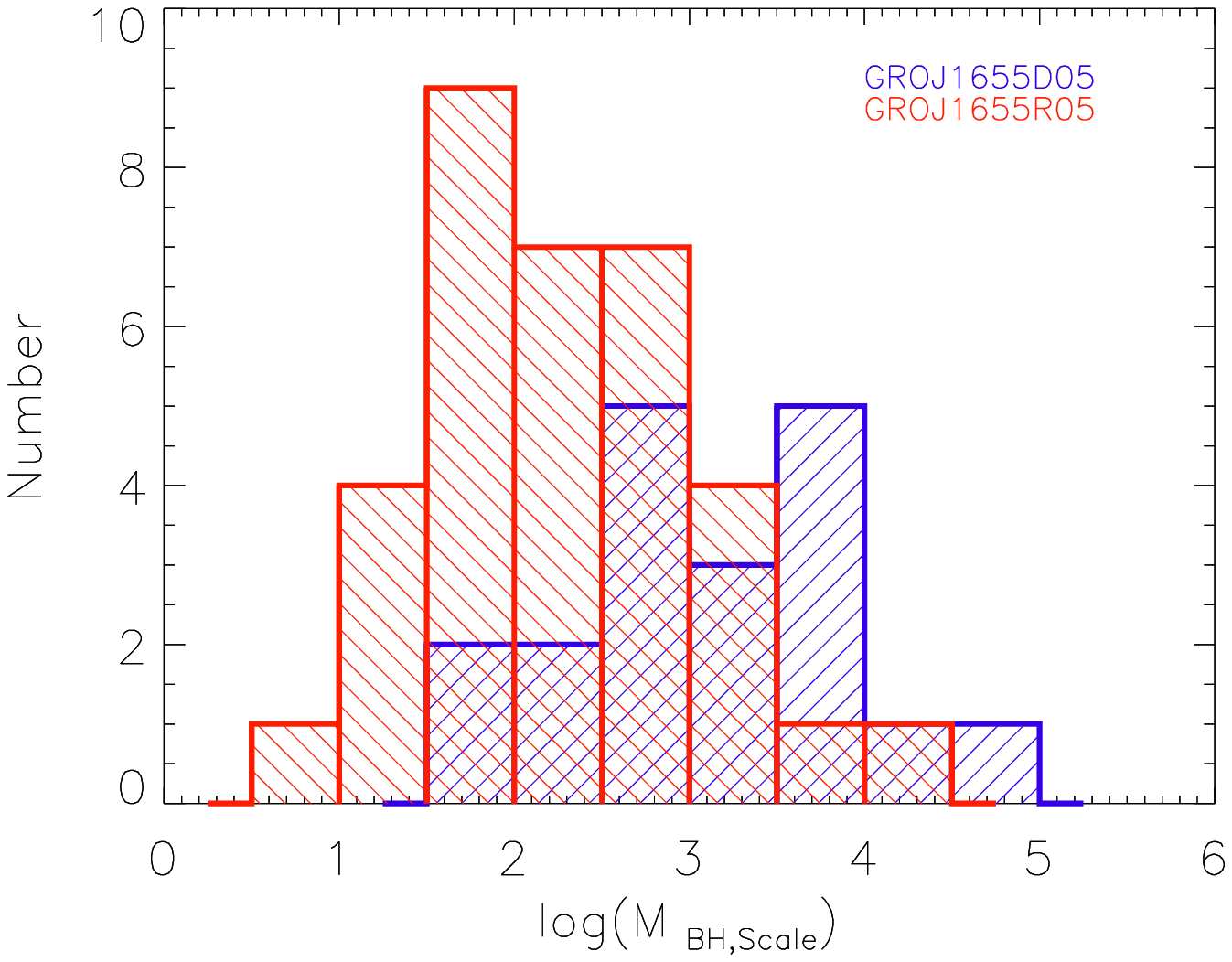}
\includegraphics[scale=0.5]{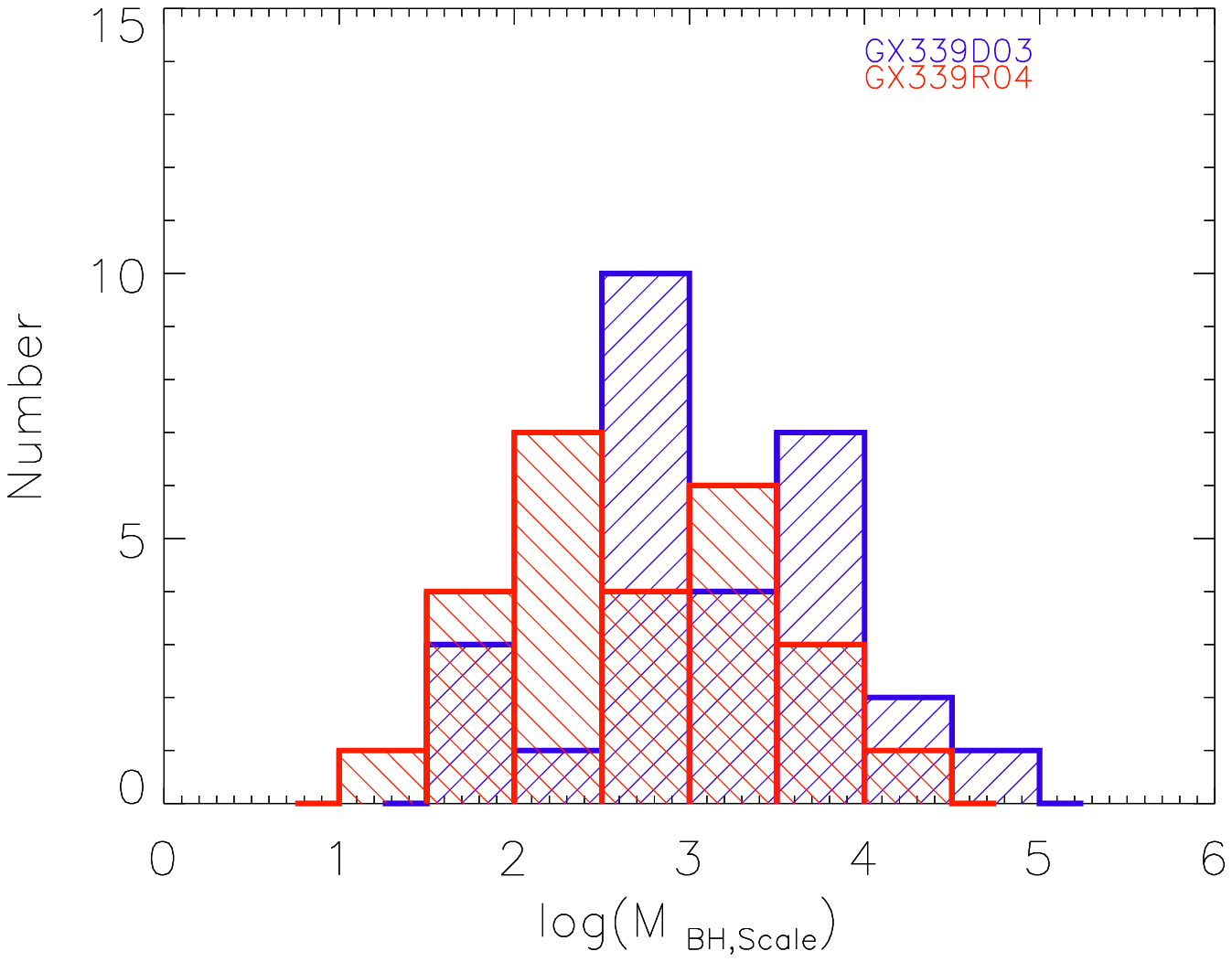}
\includegraphics[scale=0.5]{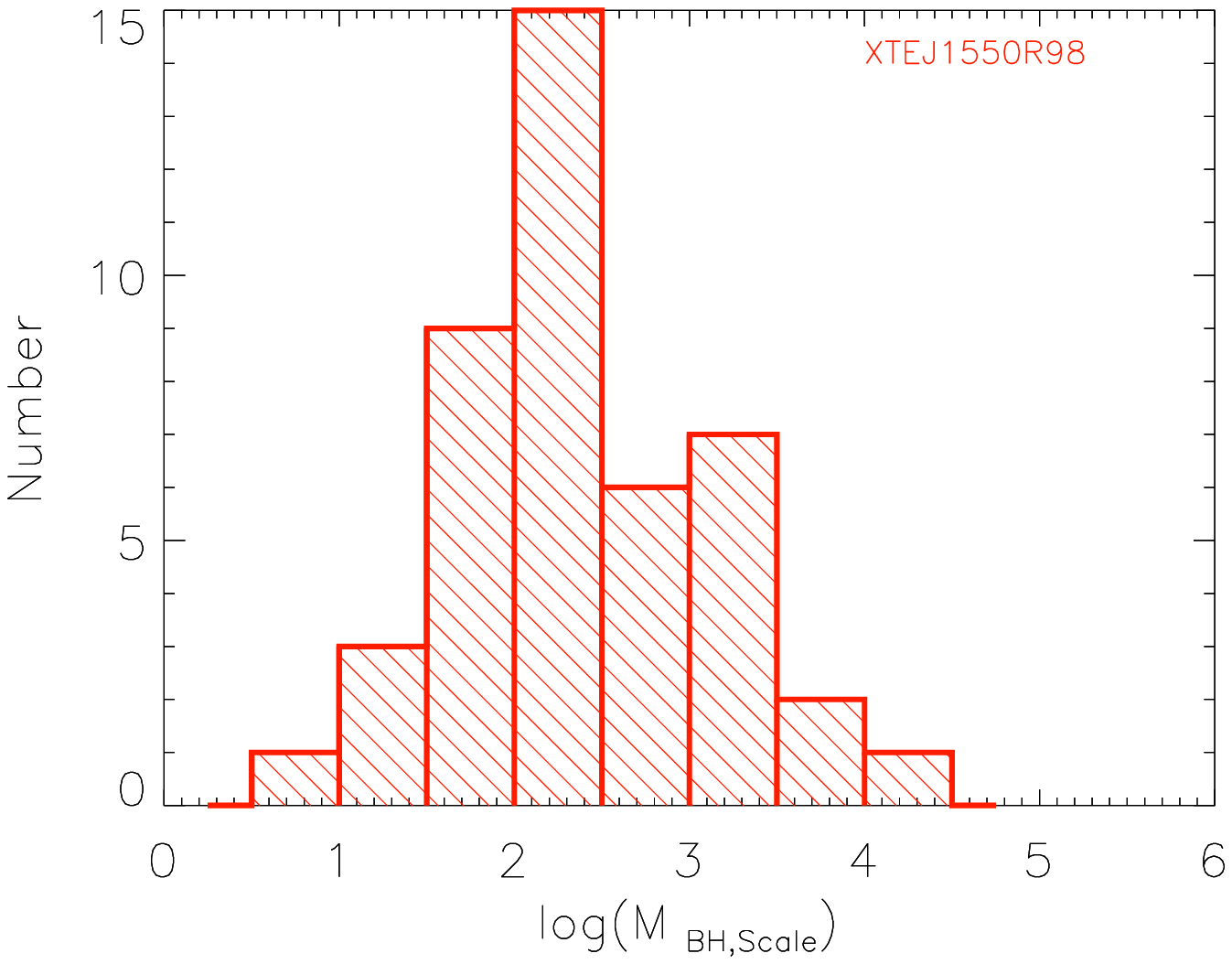}
\includegraphics[scale=0.5]{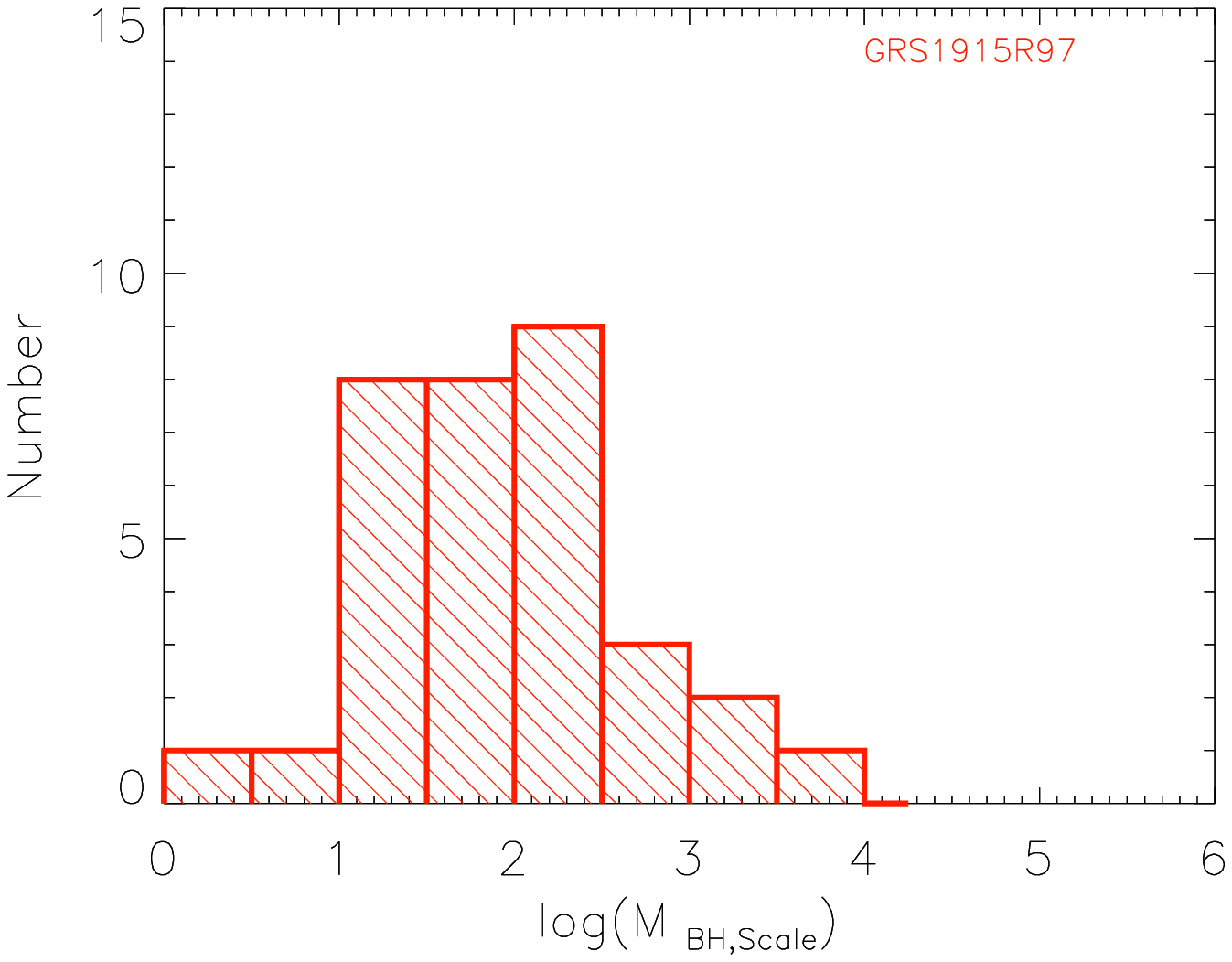}
\caption[The distribution of $M_{\rm BH,Scale}$]
{\footnotesize The distribution of $M_{\rm BH,Scale}$ the value of $M_{\rm BH}$ obtained with the X-ray scaling method. 
The $M_{\rm BH,Scale}$ histograms obtained using the decay phase are indicated by the negative slope (blue in color version), whereas the positive slope (red) histograms indicate those obtained using reference sources during the outburst rise. 
The used reference pattern is indicated at the top-right corner of each plot.}
\label{MBH_Distribution}
\end{figure*}

We then used the best-fit results to compute the black hole mass values using Eq. \ref{equ2}. 
The scaled black hole mass ($M_{\rm BH,Scale}$) values were generally distributed in the range of  $10-10^4$ \msun\ and values obtained from the decay reference episodes were generally larger by a factor of $2-3$ compared to those obtained from the rise reference episodes. 
The average of computed $M_{\rm BH,Scale}$ by the rise patterns ($<\log(M_{\rm BH,Scale})>=2.32\pm0.74$) was within $1\sigma$ from the value of decay patterns ($=3.11\pm0.76$). 
The distribution of the computed \mbh\ values for each reference pattern is illustrated in Figure \ref{MBH_Distribution}. 
Table \ref{ulxtab5} summarizes the number of ULXs with $M_{\rm BH,Scale}$ $<100$ \msun\  and $\ge100$ \msun, as well as the average of $M_{\rm BH,Scale}$ value obtained from each reference pattern. 

\begin{table} 
\begin{adjustwidth}{-0.8 cm}{}
\footnotesize
\caption{The summary of X-ray scaling method results}
\label{ulxtab5}
\begin{center}
\begin{tabular}{lccc} 
\hline        
\hline
\noalign{\smallskip}
\multicolumn{1}{l}{Reference} & \multicolumn{2}{c}{Number of ULXs} & $<\log(M_{\rm BH,Scale})>$\\
\multicolumn{1}{l}{pattern}  & $<100$ \msun\ & $\ge100$ \msun\ &\\
\noalign{\smallskip}
\hline
\hline
\noalign{\smallskip}
GROJ1655D05 & 2 & 17 & $3.12\pm0.76$\\
GROJ1655R05 & 14 & 19 & $2.31\pm0.73$\\
GX339D03 & 3 & 25 & $3.09\pm0.76$\\
GX339R04 & 5 & 21 & $2.64\pm0.77$\\
XTE1550R98 & 13 & 30 & $2.38\pm0.73$\\
GRS1915R97 & 18 & 15 & $1.95\pm0.72$\\
\hline
\hline
\end{tabular}
\end{center}
\end{adjustwidth}
\end{table}

The detailed results of this analysis are reported in Table \ref{MBHResults} where 
column (1) is the host galaxy, (2) ULX, columns from (3) to (8) represent the 
computed $M_{\rm BH,Scale}$ values using the different reference 
patterns (GROJ1655D05, GROJ1655R05, GX339D03, GX339R04, 
XTEJ1550R98, and GRS1915R97) in logarithmic scale, column (9) the value 
of $M_{\rm BH}$ reported in the literature ($M_{\rm BH,Lit}$), column (10) the 
corresponding reference of $M_{\rm BH,Lit}$. 

\begin{table*} 
\scriptsize
\caption{Estimated Mass of BH in ULX}
\begin{center}
\scriptsize
\begin{tabular}{llccccccll} 
\hline        
\hline
\noalign{\smallskip}
\noalign{\smallskip}
\multicolumn{1}{c}{Galaxy} & \multicolumn{1}{c}{ULX} & GROJ1665D05 & GROJ1655R05 & GX3994D03 & 
GX399R04 & XTEJ15550R98 & GRSJ1995R97 & $M_{\rm BH,Lit}$ & Reference\\
\multicolumn{1}{c}{(1)} & \multicolumn{1}{c}{(2)} & (3) & (4) & (5) & (6) & (7) & (8) & \multicolumn{1}{c}{(9)} & 
\multicolumn{1}{c}{(10)}\\
\noalign{\smallskip}
\hline
\noalign{\smallskip}
HoII&X-1&$ \cdots$&$\cdots$&$\cdots$&$\cdots$&$2.56\pm0.60$&$2.41\pm0.38$&$2.15\pm0.85$&$1-5$\\
HoIX&X-1	&$4.12\pm0.01$&$3.56\pm0.01$&$4.32\pm0.13$&$3.91\pm0.16$&$3.73\pm0.31$&$3.11\pm0.19$&$2.76\pm1.06$&$2-9$\\
IC 342&X-1&$3.57\pm0.04$&$3.19\pm0.17$&$3.91\pm0.18$&$3.34\pm0.37$&$3.16\pm0.22$&$2.74\pm0.02$&$3.50\pm0.97$&3,4,9\\
&XMM2&$3.72\pm0.04$&$3.31\pm0.01$&$4.13\pm0.29$&$3.67\pm0.37$&$3.38\pm0.33$&$3.10\pm0.02$&$\ge3.18$&4\\
&XMM3&$\cdots$&$\cdots$&$\cdots$&$\cdots$&$1.85\pm0.16$&$1.55\pm0.02$&$\ge2.87$&4\\
&XMM4&$\cdots$&$2.00\pm0.12$&$\cdots$&$\cdots$&$1.95\pm0.16$&$1.46\pm0.02$&$\cdots$&\\
M31&ULX& $2.66\pm0.34$&$1.94\pm0.19$&$2.51\pm0.36$&$2.01\pm0.56$&$1.71\pm0.48$&$1.11\pm0.48$&$\cdots$&\\
M33&X-8& $\cdots$&$1.97\pm0.17$&$0.00\pm0.00$&$\cdots$&$2.08\pm0.16$&$1.82\pm0.20$&$2.09\pm1.09$&10,11\\
M81&X-6& $2.40\pm0.42$&$1.90\pm0.37$&$2.62\pm0.37$&$2.09\pm0.42$&$2.09\pm0.74$&$0.82\pm0.62$&$1.16\pm0.77$&2,4,12\\
M82&X-1& $4.95\pm0.03$&$4.25\pm0.09$&$4.95\pm0.13$&$4.30\pm0.34$&$4.10\pm0.56$&$3.75\pm0.06$&$4.36\pm0.28$&$13-16$\\
NGC 1313&X-1& $3.70\pm0.09$&$3.14\pm0.27$&$3.66\pm0.08$&$3.52\pm0.13$&$3.18\pm0.21$&$2.92\pm0.45$&$3.29\pm0.51$& 2,3,4,6,9,17,28\\
& X-2& $3.60\pm0.37$&$2.79\pm0.62$&$3.76\pm0.37$&$3.25\pm0.58$&$3.07\pm0.76$&$2.14\pm0.21$&$2.96\pm0.15$& 2,3,4,6,9,17\\
& XMM2& $\cdots$&$\cdots$&$\cdots$&$\cdots$&$2.11\pm0.39$&$2.26\pm0.17$&$\cdots$&\\
& XMM4& $2.94\pm0.28$&$2.42\pm0.33$&$3.14\pm0.33$&$2.57\pm0.21$&$2.48\pm0.41$&$\cdots$&$\ge2.06$&4\\
NGC 2403&X-1& $\cdots$&$\cdots$&$\cdots$&$\cdots$&$3.30\pm0.16$&$\cdots$&$1.35\pm0.12$&2,4\\
NGC 253&X-1& $2.41\pm0.04$&$1.93\pm0.08$&$2.63\pm0.06$&$1.98\pm0.12$&$1.86\pm0.29$&$1.50\pm0.11$&$1.18\pm0.79$&3,4,12,18,19\\
& X-2& $3.22\pm0.04$&$2.36\pm0.40$&$3.34\pm0.14$&$\cdots$&$2.40\pm0.19$&$1.92\pm0.12$&$1.49\pm0.49$&3,4,12\\
& XMM4& $\cdots$&$\cdots$&$\cdots$&$\cdots$&$2.21\pm0.38$&$\cdots$&$\cdots$&\\
& XMM5& $2.80\pm0.20$&$2.02\pm0.12$&$2.75\pm0.13$&$2.27\pm0.15$&$2.06\pm0.17$&$1.59\pm0.14$&$\cdots$&\\
NGC 300&XMM1& $1.79\pm0.04$&$1.20\pm0.01$&$1.99\pm0.08$&$1.51\pm0.24$&$1.28\pm0.16$&$1.10\pm0.01$&$1.43\pm0.17$&20\\
& XMM2& $\cdots$&$0.83\pm0.17$&$1.51\pm0.21$&$1.14\pm0.24$&$0.74\pm0.16$&$0.30\pm0.06$&$\cdots$&\\
& XMM3& $1.72\pm0.04$&$1.11\pm0.10$&$1.85\pm0.10$&$1.52\pm0.06$&$1.16\pm0.08$&$\cdots$&$\cdots$&\\
NGC 4395&XMM1& $\cdots$&$1.43\pm0.35$&$\cdots$&$\cdots$&$1.19\pm0.16$&$1.33\pm0.02$&$\ge1.36$&4\\
& XMM2& $\cdots$&$1.49\pm0.30$&$2.22\pm0.29$&$1.83\pm0.18$&$1.50\pm0.24$&$\cdots$&$\cdots$&\\
& XMM3& $\cdots$&$1.92\pm0.29$&$2.68\pm0.31$&$2.10\pm0.06$&$1.80\pm0.12$&$\cdots$&$\cdots$&\\
NGC 4490&XMM1&$3.32\pm0.15$&$2.79\pm0.11$&$3.52\pm0.10$&$3.17\pm0.06$&$2.83\pm0.09$&$2.22\pm0.08$&$1.00\pm0.60$&4,26\\
& XMM2& $\cdots$&$\cdots$&$\cdots$&$\cdots$&$3.78\pm0.16$&$\cdots$&$0.96\pm0.66$&4,26\\
& XMM3& $\cdots$&$\cdots$&$\cdots$&$\cdots$&$3.07\pm0.83$&$2.28\pm0.02$&$2.24\pm0.90$&4,26\\
& XMM4& $\cdots$&$2.64\pm0.17$&$\cdots$&$\cdots$&$2.31\pm0.63$&$2.27\pm0.18$&$1.41\pm0.71$&4,26\\
& XMM5& $\cdots$&$2.75\pm0.01$&$3.49\pm0.01$&$3.23\pm0.24$&$3.22\pm0.38$&$2.26\pm0.02$&$\ge2.99$&4\\
NGC 4736&XMM1&$\cdots$&$1.93\pm0.08$&$2.72\pm0.03$&$2.42\pm0.24$&$2.04\pm0.16$&$1.46\pm0.02$&$\ge2.32$&4\\
NGC 4945&XMM1&$\cdots$&$2.16\pm0.11$&$2.94\pm0.06$&$2.52\pm0.12$&$2.19\pm0.16$&$1.67\pm0.02$&$\cdots$&\\
& XMM2& $\cdots$&$\cdots$&$\cdots$&$\cdots$&$2.49\pm0.43$&$\cdots$&$\cdots$&\\
NGC 5194&XMM1&$\cdots$&$1.86\pm0.37$&$2.78\pm0.19$&$2.53\pm0.16$&$2.00\pm0.30$&$1.49\pm0.23$&$\ge3.86$&4,21\\
& XMM2& $2.78\pm0.67$&$2.27\pm0.73$&$3.02\pm0.75$&$2.43\pm0.52$&$2.15\pm0.57$&$\cdots$&$1.38\pm0.08$&4,21\\
& XMM3& $3.59\pm0.36$&$3.14\pm0.37$&$3.89\pm0.39$&$3.11\pm0.34$&$2.90\pm0.35$&$\cdots$&$2.81\pm0.60$&4,21\\
& XMM4& $3.14\pm0.31$&$2.68\pm0.36$&$3.50\pm0.46$&$2.67\pm0.31$&$2.34\pm0.32$&$\cdots$&$\ge2.29$&4,21\\
& XMM5& $\cdots$&$2.36\pm0.15$&$\cdots$&$\cdots$&$2.41\pm0.37$&$1.91\pm0.47$&$\cdots$&\\
& XMM6& $\cdots$&$1.67\pm0.19$&$2.51\pm0.09$&$\cdots$&$1.91\pm0.31$&$1.24\pm0.15$&$1.74\pm0.74$&4,21\\
& XMM7& $2.84\pm0.33$&$1.95\pm0.01$&$2.67\pm0.01$&$2.29\pm0.24$&$1.96\pm0.16$&$1.51\pm0.02$&$1.15\pm0.15$&4,21\\
NGC 5204&X-1& $\cdots$&$2.89\pm0.01$&$3.64\pm0.01$&$3.30\pm0.03$&$2.90\pm0.18$&$2.45\pm0.13$&$\ge2.46$&2,3,4,22\\
NGC 5408&X-1& $\cdots$&$\cdots$&$\cdots$&$\cdots$&$2.84\pm0.82$&$2.69\pm0.86$&$2.81\pm0.91$&23,24\\
M101&X-1& $\cdots$&$\cdots$&$\cdots$&$\cdots$&$1.53\pm0.44$&$1.41\pm0.14$&$1.39\pm0.09$&27\\
NGC 6949&X-1& $\cdots$&$2.75\pm0.24$&$\cdots$&$\cdots$&$2.69\pm0.29$&$2.43\pm0.13$&$\le3$&3,25\\
\noalign{\smallskip}
\hline        
\hline
\noalign{\smallskip}
\end{tabular}
\end{center}
\scriptsize
\label{MBHResults}
\begin{itemize}
\item[] All estimated values are in logarithmic scale.
\item[] {\it Reference Note} $-$ 1) Goad et al. 2005, 2) Gonz\'{a}lez$-$Mart\'{i}n et al. 2011, 
3) Kajava et al. 2009, 4) Winter et al. 2006, 5) Zampieri \& Roberts 2009, 
6) Heil et al. 2009, 7) Dewangan, Titarchuk, \& Griffiths 2006, 8) Tsunoda et al. 2006, 
9) Wang et al. 2004, 10) Foschini et al. 2004, 11) Geghardft et al. 2001, 
12) Hui \& Krolik 2008, 13) Feng, Rao, \& Kaaret 2010, 14) Kaaret et al. 2001, 
15) Yuan et al. 2007, 16) Feng \& Kaaret 2010, 17) Miller et al. 2003, 
18) Banard et al. 2010, 19) Bauer et al. 2005, 20) Carpono et al. 2007, 
21) Dewangan et al. 2005, 22) Vierdaynati et al. 2006, 
23) Soria et al. 2004, 24) Strohmayer et al. 2007, 25) Rao, Feng, \& Kaaret 2010, 
26) Yoshida et al. 2010, 27) Liu et al. 2013, 28) Pasham et al. 2015
\end{itemize}
\end{table*}


There were 5 ULXs (IC 345 XMM4, NGC 4395 XMM1, XMM2, NGC 4945 XMM2, and NGC 4490 XMM4) 
whose $\Gamma-N_{\rm BMC}$ diagrams were constructed with only two  data points where one of their 
\gam\ values was outside the range of any reference pattern or had a very large uncertainty 
(e.g., $\sigma_{\Gamma}\ge1$). 
NGC 2403 X-1 had all three measured \gam\ values below $\approx1.4$ hampering the comparison with 
any reference trend and consequently the \mbh\ computation. 
Similarly, NGC 4490 XMM3 and NGC 4490 XMM5 had 2 out of 3 meausred \gam\ values 
$\sim1$ and their $\Gamma-N_{\rm BMC}$ could not to be compared to any reference pattern. 
These 8 ULXs were excluded from further analysis. 

M101 ULX-1 is the only ULX for which \mbh\ has been obtained dynamically \citep{liu13}. 
Applying our systematic procedure to the three \xmm\ observations available for this source, we obtained a value of $M_{\rm BH}=19-35\,M_{\odot}$ which is consistent with the range measured dynamically. 
However, Titarchuck and Seifina (2016), using {\it Chandra} and {\it Swift} data and the {\it BMC} model derived a $M_{\rm BH}$ value of the order of $10^4$ $M_{\odot}$, arguing that the value derived by Liu et al. (2013) is underestimated because the inclination and local absorption are not properly accounted for. 
Given the large discripancy with Titarchuk results, we performed an additional spectral analysis of the {\it Chandra} and \xmm\ data-sets of M101 ULX-1. 
Using same model of Titarchuk and Sefina (2016), phabs*bmc and restricting the fitting range of $0.3-7$ keV (as opposed to our standard procedure which fits the $0.3-10$ keV range with the baseline model wabs*(diskpn+bmc)), we were able to replicate Titarchuk and Sefina (2016) results only when $n_H$ was fixed at the value of $3\times10^{21}$ cm$^2$.
We conclude that the value of $M_{\rm BH}$ derived for M101 ULX-1 should be taken with caution, given the ongoing debate and uncertainties related to the srource's intrinsic absorption. 

Recent results indicate that the compact objects in some ULXs are actually neutron stars. For example M82 X-2, NGC 5907 ULX, and NGC 7793 P13 \citep{israel16, israel17a, israel17b}. For completeness, we tried to extend the X-ray scaling method to these sources. However, this method cannot be applied to NGC 7793 P13, because there are only two XMM-Newton archival observations, of which one has a spectrum with $\Gamma$ flatter than 1.3, which cannot be compared with any of the reference sources.  M82 X-2 has more than 15 XMM-Newton observations, but the limited spatial resolution of the EPIC cameras does not allow one to disentangle the emission associated with the ULX from that of the host galaxy and hence to determine the black hole mass with the X-ray scaling method. Only NGC 5907 ULX has XMM-Newton data that can be used to estimate the mass of the compact object using the X-ray scaling method. We carried out the same procedure described in previous sections; of the eight spectra analyzed two yield photon indices flatter than 1.2 and hence are excluded from further analysis. The remaining six observations plotted in the $\Gamma-N_{\rm BMC}$ diagram do not show a clear positive correlation, suggesting that the spectral evolution of this source is different from the typical trend shown by the reference BH sources and by the majority of the objects analyzed in this work. Nevertheless, we applied the scaling method and obtained a black hole mass of the order of 1000 solar masses. This result is not surprising, since the implicit assumption of this method are that the bulk of the X-ray radiation is the quasi-isotropic emission from the corona, and that all black holes show a similar spectral evolution irrespective of their mass. If one of these conditions is not fulfilled (as in the case of highly beamed X-ray emission,  anomalous spectral transition, or X-rays produced by a neutron star), the resulting mass of the compact object will be overestimated.

\subsection{Correlation Analysis}
\subsubsection{Comparison with \mbh\ from different methods}
We looked for correlations between $M_{\rm BH,Scale}$ values and the corresponding values reported in the literature. 
The values of $M_{\rm BH}$ quoted in the literature for ULXs are based on different methods. 
Some are obtained using the relationship between the mass and luminosity, others using the inverse correlation between the mass and QPO frequency. 
As a result, \mbhlit\ values for the same source may span a wide range (sometimes with a few orders of magnitude difference). 
Therefore, we made correlation studies of $M_{\rm BH,Scale}$ values with the minimum and maximum \mbhlit\ values ($M_{\rm BH,Lit,Min}$ and $M_{\rm BH,Lit,Max}$, respectively) and also with the mean of \mbhlit\ values ($M_{\rm BH,Lit,Mean}$). 

\begin{figure*}
\includegraphics[scale=0.5]{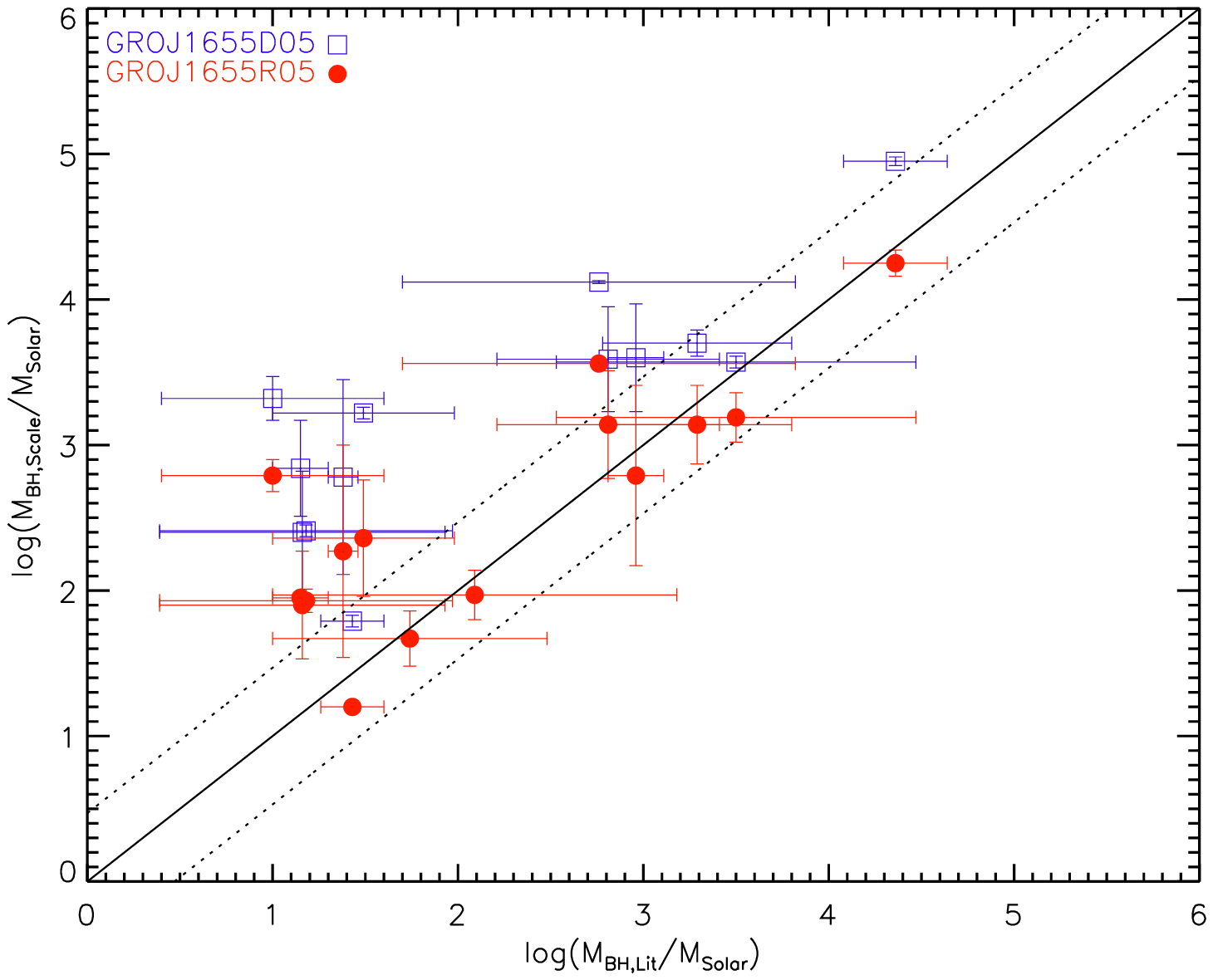}
\includegraphics[scale=0.5]{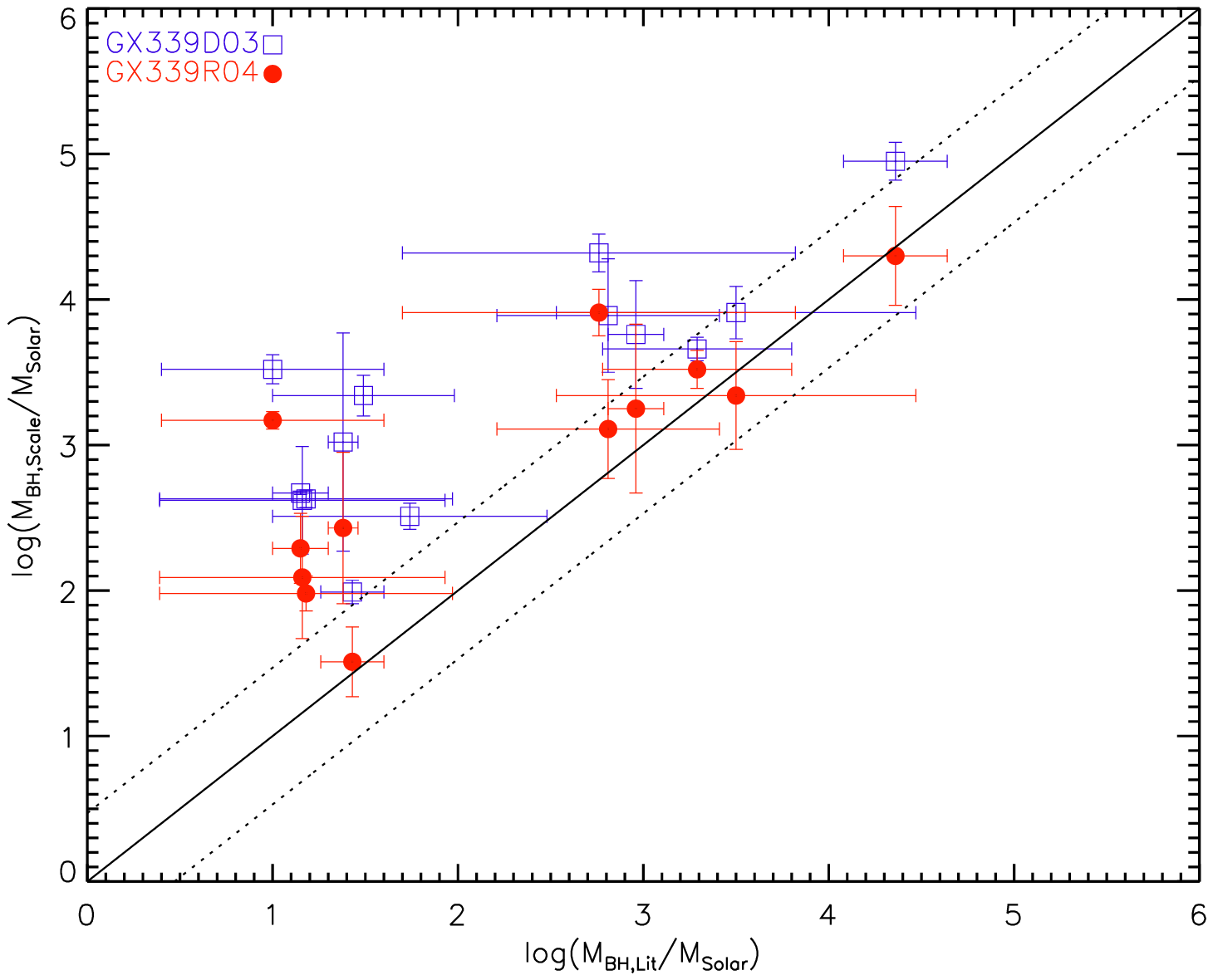}\\
\includegraphics[scale=0.5]{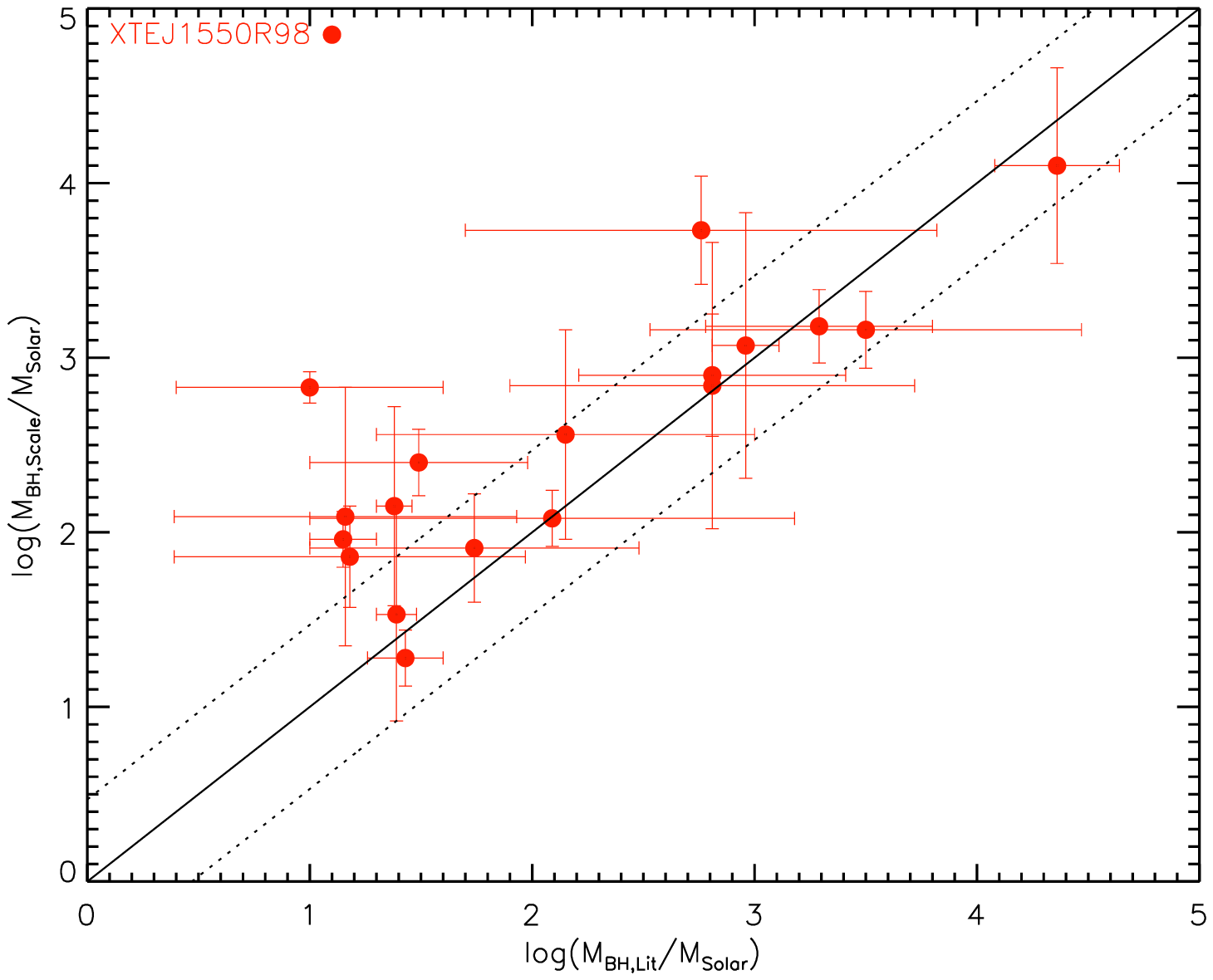}
\includegraphics[scale=0.5]{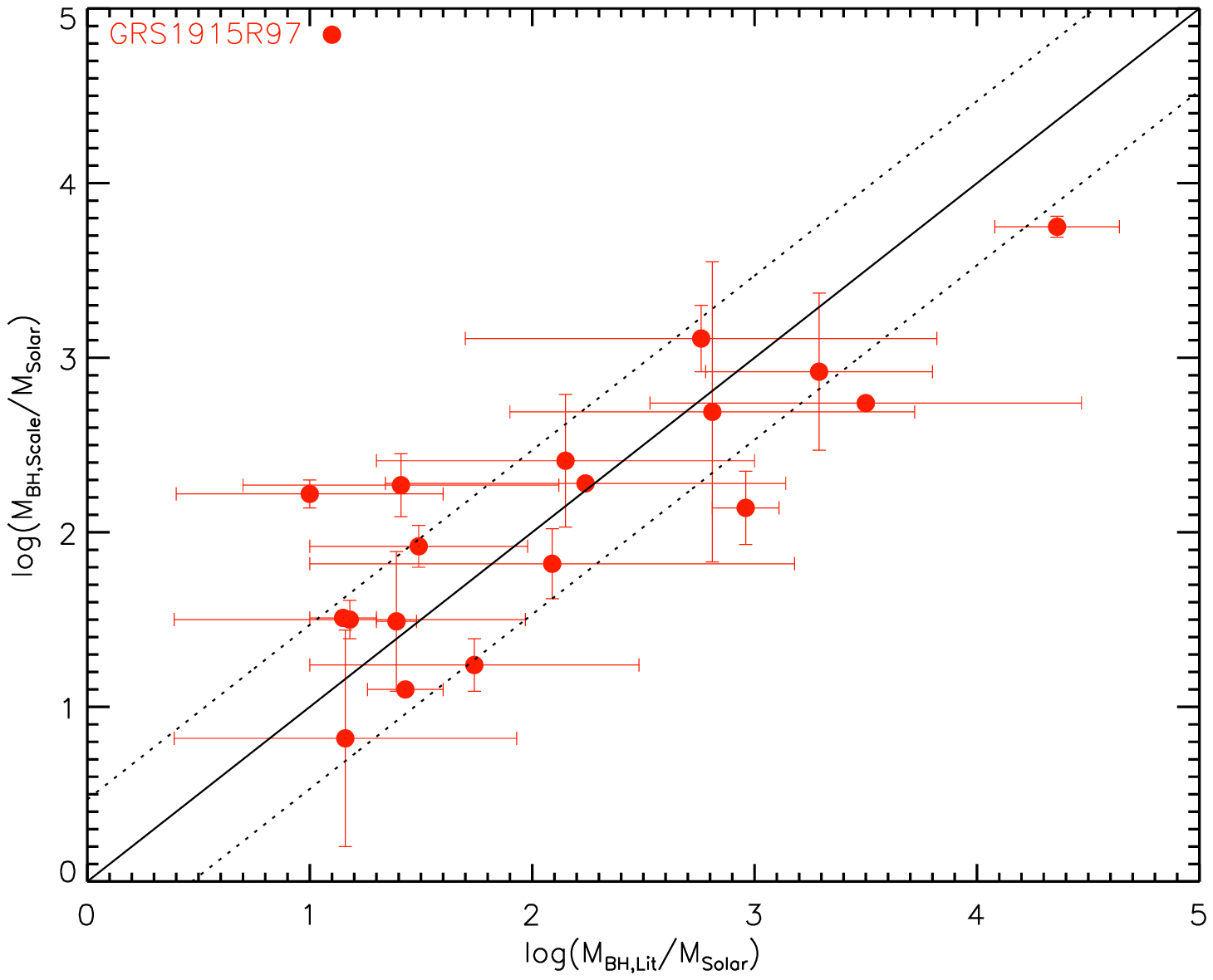}\\
\caption[Plot of $M_{\rm BH,Scale}$ vs. \mbhlit\ for all ULXs]
{\footnotesize Plot of $M_{\rm BH,Scale}$ vs. \mbhlit\ for all ULXs. We used filled circles to 
indicate $M_{\rm BH,Scale}$ values from rising patterns and open squares for decay patterns. 
The solid line indicates the one-to-one correlation between $M_{\rm BH,Scale}$ and \mbhlit\ 
and the dash lines for the 0.47 dex boundaries.}
\label{MBHcorrelation}
\end{figure*}

We compared $M_{\rm BH,Scale}$ to the corresponding $M_{\rm BH,Lit,Mean}$ for every pattern. 
The linear correlation results suggest that \mbh\ values for all reference patterns were broadly consistent with the corresponding $M_{\rm BH,Lit,Mean}$ within $1-2\sigma$ uncertainty. 
The linear correlation results of $M_{\rm BH,Scale}$ value from corresponding the $M_{\rm BH,Lit,Min}$, $M_{\rm BH,Lit,Max}$, and $M_{\rm BH,Lit,Mean}$ values are reported in Table \ref{MBHCorrelation} for each reference pattern with the best-fit slope, the intercept, Spearman's $\rho-$rank and its following probability, and the RMS value. 
We used the {\small MPFITEXY} routine \citep{markwardt2009, williams2010} which accounts for errors on both axes for the comparisons. 
We also plotted $\log(M_{\rm BH,Scale})$ versus $\log(M_{\rm BH,Lit,Mean})$ in Figure \ref{MBHcorrelation} for each reference pattern. 
The visual inspection of these plots confirms that the X-ray scaling method estimates of \mbh\ using the rising patterns are in good agreement with $M_{\rm BH,Lit,Mean}$ in both sMBHs and IMBHs. 

\begin{table}
\begin{adjustwidth}{-1.5 cm}{}
\footnotesize
\caption{\mbh\ correlation analysis}
\label{MBHCorrelation}
\begin{center}
\begin{tabular}{lcccc} 
\hline        
\hline
\noalign{\smallskip}
\multicolumn{1}{c}{Reference pattern} & Slope & Intercept & Spearman & RMS\\
\multicolumn{1}{c}{(1)} & (2) & (3) & (4) & (5)\\
\noalign{\smallskip}
\hline
\noalign{\smallskip}
\multicolumn{5}{c}{$\log(M_{\rm BH,Scale})$ versus $\log(M_{\rm BH,Lit,Mean})$}\\
\hline
GROJ1655D05 & $1.03\pm 0.10$ & $0.71\pm 0.24$ & 0.71($6.7\times10^{-3}$) & 1.23\\
GROJ1655R05 & $ 1.02\pm 0.10$ & $ 0.29\pm 0.24$ & 0.67($5.8\times10^{-3}$) & 0.71\\
GX339D03 & $ 0.94\pm 0.10$ & $ 1.22\pm 0.23$ & 0.67($8.1\times10^{-3}$) & 1.33\\
GX339R04 & $ 0.83\pm 0.12$ & $ 0.97\pm 0.30$ & 0.67($1.7\times10^{-2}$) & 0.92\\
XTEJ1550R98 & $ 0.95\pm 0.15$ & $ 0.49\pm 0.34$ & 0.73($6.0\times10^{-4}$) & 0.67\\
GRS1915R97 & $ 0.82\pm 0.07$ & $ 0.26\pm 0.17$ & 0.73($5.3\times10^{-4}$) & 0.54\\
\hline
\noalign{\smallskip}
\multicolumn{5}{c}{$\log(M_{\rm BH,Scale})$ versus $\log(M_{\rm BH,Lit,Min})$}\\
\hline
GROJ1655D05 & $0.65\pm0.13$ & $1.95\pm0.32$ &0.75($8.4\times10^{-4}$) & 1.59\\
GROJ1655R05 & $0.71\pm 0.13$ & $1.08\pm0.30$ & 0.66($8.4\times10^{-4}$) &0.98\\
GX339D03 & $0.66\pm0.13$ & $2.01\pm0.31$ & 0.72($4.6\times10^{-4}$) &1.67\\
GX339R04 & $0.69\pm0.15$ & $1.47\pm0.35$ & 0.75($5.6\times10^{-4}$) &1.20\\
XTEJ1550R98 & $0.56\pm0.14$ & $1.33\pm0.31$ & 0.55($3.3\times10^{-3}$) &1.00\\
GRS1915R97 & $0.60\pm0.13$ & $0.84\pm0.30$ & 0.63($1.7\times10^{-3}$) &0.79\\
\hline
\noalign{\smallskip}
\multicolumn{5}{c}{$\log(M_{\rm BH,Scale})$ versus $\log(M_{\rm BH,Lit,Max})$}\\
\hline
GROJ1655D05 & $ 0.83\pm 0.11$ & $ 1.06\pm 0.31$ &0.82($8.8\times10^{-5}$) & 0.87\\
GROJ1655R05 & $ 0.84\pm 0.11$ & $ 0.42\pm 0.28$ &0.76($3.5\times10^{-5}$) &0.59\\
GX339D03 & $ 0.79\pm 0.12$ & $ 1.31\pm 0.30$ &0.79($6.6\times10^{-5}$) & 0.96\\
GX339R04 & $ 0.77\pm 0.13$ & $ 0.99\pm 0.34$ &0.82($6.5\times10^{-5}$) & 0.66\\
XTEJ1550R98 & $ 0.76\pm 0.12$ & $ 0.50\pm 0.31$ &0.76($6.4\times10^{-6}$) & 0.60\\
GRS1915R97 & $ 0.74\pm 0.11$ & $ 0.16\pm 0.29$ & 0.80($7.2\times10^{-6}$) & 0.83\\
\hline
\hline
\end{tabular}
\end{center}
\end{adjustwidth}
{\footnotesize {\it Note.} Column (1) a reference pattern; (2) a best-fit slope; (3) a best-fit 
intercept; (4) Spearnan's $\rho-$rank and its following probability; (5) RMS value from 
the one-to-one correlation}
\end{table}

The correlation study between $M_{\rm BH,Scale}$ and \mbhlit\ values can be summarized as following. 
All the patterns showed a general agreement, suggesting that different reference patterns provide consistent values of \mbh.
For all ULXs, we compared the $M_{\rm BH,Scale}$ values obtained from different reference patterns with the corresponding $M_{\rm BH,Scale}$ values based on the XTEJ1550R98 pattern (hereafter, $M_{\rm BH,XTE}$).
We used XTEJ1550R98 as primary reference because 1) its spectral pattern in the $\Gamma-N_{\rm BMC}$ diagram spans the largest range of \gam\ allowing the determination of \mbh\ for the vast majority of the ULXs in our sample and 2) it provided the best agreement with the \mbh\ values obtained with different methods. 

The $M_{\rm BH}$ values obtained with the XTEJ1550R98 pattern are fully consistent with those obtained with all the other reference patterns, as demonstrated by the results of a linear correlation analysis summarized in Table \ref{MBHXTECorrelation}. 

\begin{table}
\begin{adjustwidth}{-1.4 cm}{}
\footnotesize
\caption{$M_{\rm BH,Scale} - M_{\rm BH,XTE}$ correlation analysis}
\label{MBHXTECorrelation}
\begin{center}
\begin{tabular}{lcccc} 
\hline        
\hline
\noalign{\smallskip}
Ref. pattern & Slope & Intercept & Spearman & RMS\\
\multicolumn{1}{c}{(1)} & (2) & (3) & (4) & (5)\\
\noalign{\smallskip}
\hline
GROJ1655D05 & $ 1.01\pm 0.06$ & $-0.58\pm 0.21$ & 0.96($8.5\times10^{-11}$) & 0.63\\
GROJ1655R05 & $ 1.00\pm 0.06$ & $ 0.03\pm 0.16$ & 0.95($5.2\times10^{-16}$) & 0.16\\
GX339D03 & $ 1.01\pm 0.07$ & $-0.74\pm 0.21$ & 0.95($2.0\times10^{-14}$) & 0.75\\
GX339R04 & $ 1.02\pm 0.06$ & $-0.36\pm 0.17$ & 0.96($2.5\times10^{-14}$) &0.29\\
GRS1915R97 & $ 1.04\pm 0.07$ & $ 0.37\pm 0.15$ & 0.91($5.3\times10^{-12}$) &0.54\\
\hline
\hline
\end{tabular}
\end{center}
\end{adjustwidth}
{\footnotesize {\it Note.} Column (1) a reference pattern; (2) a best-fit slope; (3) a best-fit 
intercept; (4) Spearnan's $\rho-$rank and its following probability; (5) RMS value from 
the one-to-one correlation}
\end{table}
\subsubsection{Comparison between $M_{\rm BH,Scale}$ and $M_{\rm BH,QPO}$}
\begin{figure} 
\includegraphics[scale=0.5]{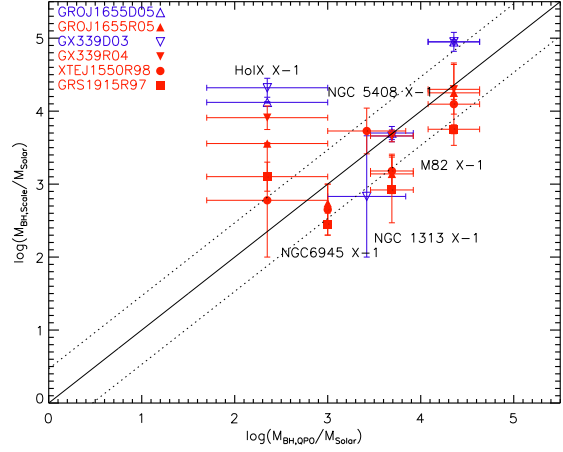}
\caption[Plot of $M_{\rm BH,Scale}$ versus $M_{\rm BH,QPO}$.]
{\footnotesize Plot of $M_{\rm BH,Scale}$ versus $M_{\rm BH,QPO}$. 
The plot contains $M_{\rm BH,Scale}$ values along the $y$-axis and 
$M_{\rm BH,QPO}$ values along the $x$-axis where open symbols are used 
for values from decay patterns and filled symbols for values from rise patterns. 
The solid line indicates the one-to-one correlation and dashed lines the departure 
by a factor of 3 (0.47 dex). 
The name of patterns and the corresponding used symbols are indicated at the top left 
corner.}
\label{ulxfig49}
\end{figure}

Among the methods used to constrain \mbh\ in ULXs, the technique based on QPOs 
is considered the most reliable, since unlike spectral-based methods, it is model 
independent. 
For this reason, we compared our computed $M_{\rm BH,Scale}$ values with those 
obtained via QPOs for the subsample of ULXs for which QPOs were clearly detected.  
However, we were able to test only five ULXs (HoIX X-1, NGC 1313 X-1, NGC 5408 X-1, M82 X-1, and 
NGC 6946 X-1) because the QPO based \mbh\ determination for ULXs so far very 
limited, since secure QPO detections in ULXs are elusive. 
The main finding from this comparison is that $M_{\rm BH,Scale}$ values based on 
different reference patterns (especially those associated with the rising phase of the 
outburst) show a general good agreement with the QPO based \mbh\ values. 
This comparison is illustrated in Figure \ref{ulxfig49} where we plot $\log(M_{\rm BH})$ 
vs. $\log(M_{\rm BH,QPO})$. 


\section{Summary and Conclusion}
The nature of ULXs is one of the current mysteries in high-energy astrophysics. 
They might be stellar mass BHs in a particularly bright state that can be explained by a combination of super-Eddington accretion and beaming effects. 
This appears to be the favorite interpretation for most of the ULXs with $L_{\rm X}\sim10^{39}$ \ergs, because the formation process of sMBHs is well understood and sMBHs are routinely observed in the Milky Way and nearby galaxies. 
In this framework, what is not completely understood is why this putative ultraluminous spectral state is not regularly observed in X-ray binaries in our Galaxy. 
There is however a claim that XTE J1550$-$564 went into this spectral state during the 1998 outburst. 

An alternative, perhaps more exciting, interpretation is that ULXs (at least the brightest ones, with $L_{\rm X}\sim10^{40}-10^{41}$ \ergs) host IMBHs and accrete at a regular level. 
In this case the formation process is under debate (direct collapse vs. BH mergers) but the spectral state would be consistent with the canonical ones regularly observed in GBHs. 
Finally, one cannot exclude a third intermediate possibility that ULXs are massive stellar BHs ($M_{\rm BH}\sim100\,M_{\odot}$) that accrete at high but not extreme level. 
The formation of these massive BHs can still be explained by the regular stellar evolution process under the assumption of low metallicity ($\ll1\%$ of the solar value), which should be typical for primordial stars of Population III.

These hypotheses on the nature of ULXs are not mutually exclusive (it is entirely possible that ULXs encompass sMBHs in ultraluminous states as well as highly-accreting massive stellar BHs, and normally-accreting IMBHs) and none can be ruled out until the \mbh\ is dynamically determined. 

For the time being, we need to rely on indirect methods to constrain \mbh\ in ULXs. 
With our work, we have applied the X-ray scaling method to a sample of ULXs with multiple X-ray observations. 
As explained before, this method was introduced to determine \mbh\ and distance in GBHs by scaling the X-ray spectral and temporal trend of a reference BH source whose properties were well constrained. 
We then extended this method to AGNs using the RM sample with the reasonable assumption that AGNs follow the same spectral transition as GBHs but on much longer timescales. 
Our choice of a sample of ULXs with multiple observations made it possible to compare the ULX spectral evolution with the most appropriate reference pattern. 

We performed a homogenous spectral analysis of all the available data with sufficient signal-to-noise ratio (exposure $\ge10$ ks) and then a systematic comparison of the spectral trends in the $\Gamma-N_{\rm BMC}$ plot. 
The majority of the spectral patterns show a positive trend, which can be directly compared to the reference ones. 
Some spectral trends appear more complex and can be explained by the presence of statistical outliers or by the fact that the trend may comprise data from different outbursts and/or different outburst phases (typically, the decay spectral pattern is different from the rising one). 
We cannot rule out that some spectral trends are genuinely different; in that case, it would not be possible to use the reference spectral trends to determine \mbh. 

The results of our analysis suggests that a substantial fraction of our sample is consistent with the intermediate mass BH hypothesis. 
At first sight, these findings seem to be at odds with several recent results in this field pointing out that the vast majority of ULXs are ``normal" or massive stellar BHs accreting at super-Eddington level with only few strong candidates to be intermediate mass BHs. 
However, it must be kept in mind that our sample is not complete by any means nor can be considered as representative for the whole ULX population. 
Indeed, the selection of sources with multiple and good-quality X-ray data is likely to be biased toward the brightest tail of the ULX population, which is more likely to contain larger mass objects. 
Additionally, taking into account the uncertainties associated with the \mbh\ determination, (which depend on the errors of \gam\ and $N_{\rm BMC}$ as well as on the uncertainty associated with the fitting procedure of the spectral trend in the $\Gamma-N_{\rm BMC}$ plot) the majority of the sources appear to be consistent with the hypothesis of massive sMBHs ($M_{\rm BH}\sim10^2\,M_{\odot}$) accreting at high rate. 

The fact that the \mbh\ estimated with the scaling method are largely consistent with the values obtained utilizing very different methods including variability-based methods that are model independent, seems to confirm the validity of this X-ray method at all BH scales. 
Indeed, since it has been demonstrated that the scaling method can be successfully used to constrain \mbh\ for stellar and supermassive BHs, it is natural to expect that it can also be used in the intermediate range. 
One may question the applicability of this method to ULXs by claiming that they are in a peculiar ultraluminous spectral state that cannot be compared with the standard reference patterns. 
However, we must point out that among our reference patterns we use XTE J1550$-$564, which has been identified as potential Galactic analog of ULXs in ultraluminous state. 
We also use the pattern of the historical superluminal source GRS 1915$+$105, which is known to accrete at super-Eddington rate. 
Finally, this method has been successfully used to constrain the \mbh\ of PKS 0558$-$504, a bright radio-loud Narrow Line Seyfert 1 galaxy that accretes at super-Eddington level \citep{gliozzi2010}. We therefore conclude that the scaling method can be safely used also for highly accreting objects and hence to constrain \mbh\ in ULXs. 

\begin{figure}
\footnotesize
\includegraphics[scale=0.5]{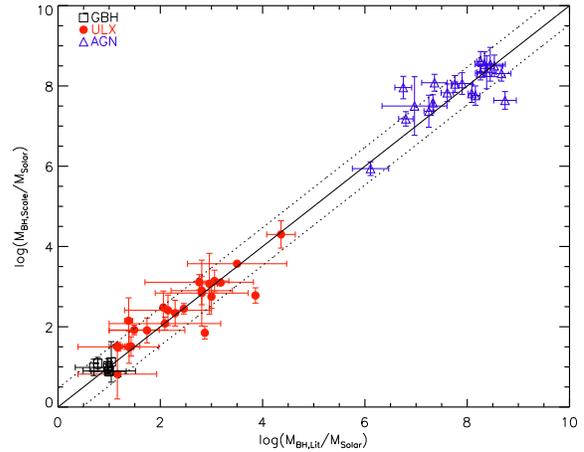}
\footnotesize
\caption[Plot of $M_{\rm BH,Scale}$ vs. $M_{\rm BH,Lit}$ for all scales]
{\footnotesize Plot of $M_{\rm BH,Scale}$ vs. $M_{\rm BH,Lit}$ for all scales. 
The obtained $M_{\rm BH}$ values from the scaling methods for GBHs are indicated 
with open squares, ULXs with filled circles, and AGNs with open triangles. The solid line 
indicates the one-to-one correlation and dash lines for 0.47 dex level.}
\label{fig4}
\end{figure}

In Figure \ref{fig4} we plot the $\log(M_{\rm BH})$ vs. $\log(M_{\rm BH,Lit})$ for GBHs, ULXs, and bright AGNs (note that GBHs and AGNs are compared to dynamically determined $M_{\rm BH}$). 
The good agreement between these values which appears evident from the image and is formally confirmed by a statistical analysis $-$ the linear best-fit slope value of $1.00\pm0.02$ and $0.02\pm0.11$ for the intercept were found and was confirmed with the one-to-one RMS value of 0.56 and the SpearmanÕs $\rho-$rank coefficient of 0.96 with its probability of $6.2\times10^{-30}$ $-$ strengthens the conclusion that the X-ray scaling method is a truly scale-independent method that can be applied to all BH systems.

\label{lastpage}
\end{document}